\documentclass[a4paper, 10pt, onecolumn]{quantumarticle}
\pdfoutput=1

\usepackage{amsmath}
\usepackage{url}
\usepackage{braket}

\usepackage{array}   
\newcolumntype{L}{>{$}l<{$}} 

\usepackage{subcaption}

\usepackage{fancyhdr}
\usepackage{fancyvrb}
\usepackage{xcolor}
\usepackage{hyperref}
\usepackage[numbers]{natbib}
\usepackage{qcircuit}
\usepackage{mathdots}

\usepackage{graphicx}
\setkeys{Gin}{width=\linewidth,totalheight=\textheight,keepaspectratio}
\graphicspath{{figures/}}

\usepackage{amssymb, amsthm, bm}
\usepackage[ruled]{algorithm2e}

\newtheorem{theorem}{Theorem}

\newcommand{\fx}{\lfloor x \rfloor}
\newcommand{\cx}{\lceil x \rceil}

\begin{document}

\title{Quantum rounding}

\author{Rajiv Krishnakumar}
\email{rajiv.krishnakumar@gs.com}
\author{William J. Zeng}
\email{william.zeng@gs.com}
\affiliation{Goldman Sachs \& Co.}

\begin{abstract}
We introduce new rounding methods to improve the accuracy of finite precision
quantum arithmetic. These quantum rounding methods are applicable when
multiple samples are being taken from a quantum program. We show how to use
multiple samples to stochastically suppress arithmetic error from rounding. We benchmark
these methods on the multiplication of fixed-point numbers stored in quantum
registers.
We show that the gate counts and depths for multiplying to a target accuracy
can be reduced by approximately 2-3X over state of the art methods while using
roughly the same number of qubits.
\end{abstract}

\section{Introduction}
Recent efforts have been made to quantify the resources needed to run
potentially commercially relevant quantum algorithms
\cite{chakrabarti2020quantum,gheorghiu2019benchmarking,gidney2019factor,meuli2020enabling}.
These algorithms often have subroutines of arithmetic operations.
Although these types of subroutines are commonplace and easily implemented on
classical computers, they can consume significant resources on quantum
computers. This is a problem as the quantum computers of today (and of the near
tomorrow) have limited quantum memory and limited quantum coherence.
Therefore it is desirable to reduce resources and precision for
quantum arithmetic.

In this paper, we propose~\emph{quantum rounding} to increase
the precision of quantum arithmetic at a fixed quantum register size.
These quantum rounding methods are applicable when
multiple samples are being taken from a quantum program.

In Section~\ref{sec:quantum-rounding} we introduce quantum rounding and
give two quantum circuit implementations. Section~\ref{sec:qr-errors} analyzes
the errors in these rounding methods and compares their scaling against
traditional rounding approaches. We find for a quantum program in which $N$
samples are taken, the worst-case quantum rounded error is suppressed as
approximately $\mathcal{O}(1 / \sqrt{N})$. In Appendix~\ref{sec:quantum-semi}
we introduce an alternative~\emph{quantum semi-rounding} method whose error we
study in an average case analysis.

Section~\ref{sec:overview-resources} presents the resources needed to scale
quantum rounding methods. In Section~\ref{sec:benchmarks} we compare
the resources used for quantum rounding with other methods when applied to fixed-point multiplication 
with an allowed target error. Figures~\ref{fig:sizes} and~\ref{fig:samples} show that the
gate counts and depths can be reduced by approximately 2-3X over state of the
art methods while using roughly the same number of qubits.

\section{Quantum Rounding}
\label{sec:quantum-rounding}
Finite precision introduces errors. Let $\bar{x}\in\mathbb{R}$ be an exact
numerical representation. The $n$-bit fixed point representation is
\begin{align}
x=\underbrace{x_{n-1}\cdots x_{n-p}}_p.\underbrace{x_{n-p-1} \cdots x_0}_{n-p}.
\end{align}
Assume that $\bar{x}$ is rational and exactly represented by $n+m$ bits.
\begin{align}
\bar{x}=\underbrace{\bar{x}_{m+n-1}\cdots \bar{x}_{m+n-p}}_p
.\underbrace{\bar{x}_{m+n-p-1} \cdots
\bar{x}_m}_{n-p}\underbrace{\bar{x}_{m-1}...\bar{x}_0}_m.
\end{align}

We then have a few choices for how to round $\bar{x}$ to $x$. Truncating by
rounding up or down introduces an error $\epsilon_{RD}\le 1 / 2^{n-p}$.
If we instead round to the nearest $n$-bit representable number, then our error
is $\epsilon_{RN}\le 1/2^{n-p-1}$.

Recently, demand for high throughput in classical machine learning applications
has driven the study of an alternative method called stochastic
rounding~\cite{connolly2021stochastic,hopkins2020stochastic,mikaitis2020stochastic,xia2020improved}.
Here, one rounds up or down with equal
probability each time. Implementing this in practice classically requires
access to a reliable and fast source of randomness.
On a quantum computer we can do this relatively easily. Let
$\lfloor x \rfloor$ and $\lceil x \rceil$ be the rounded
down and rounded up $n$-bit representations of $\bar{x}$ respectively.
To stochastically round, we add an additional
ancilla qubit and prepare it in the $\ket{+}$ state. We then do an
controlled addition of $\epsilon_{RD}$ to the rounded down state, where the
control is on the value of the $\ket{+}$ state ancilla.
This creates the state
\begin{align}
\frac{1}{\sqrt{2}}\left(\ket{\fx,0}+\ket{(\fx+\epsilon_{RD}),1}\right)
= \frac{1}{\sqrt{2}}\left(\ket{\fx,0}+\ket{\cx,1}\right).
\end{align}
Measuring the ancilla will probabilistically round up or down with equal
probability. The ancilla can then be conditionally reset to $\ket{0}$.
In this manner any benefits from stochastic rounding can be naturally obtained
in the quantum setting.

However, we can do better by making a specific choice of a non-uniform
superposition between rounding up or down. In quantum rounding, we make use of
the extra $m$ bits to bias the rounding direction
so that the expected value of many roundings converges to the higher precision.
Specifically, define the remainder
\begin{align}
\label{eq:remainder}
r = \sum_{j=1}^m2^{-j\cdot\bar{x}_{m-j}} = \frac{\bar{x}_{m-1}..
.\bar{x}_0}{2^m}.
\end{align}
to be the last $m$ bits normalized between zero and one. In this formula we
treat $\bar{x}_{m-1}...\bar{x}_0 $ as an integer.
The remainder $r$
can be interpreted as a probability of rounding up. We then prepare the ancilla
in the superposition
\begin{align}
\label{eq:phase}
\sqrt{1-r}\ket{0}+\sqrt{r}\ket{1}
\end{align}
and perform the same controlled addition as
before. This results in
\begin{align}
\ket{x} = \sqrt{1-r}\ket{\fx,0}+ \sqrt{r}\ket{\cx,1}.
\end{align}
In this final state, the expected value of the main register is exactly $\bar{x}$.
Should this rounding circuit be executed repeatedly, as is commonly the case in
quantum programs where multiple samples are taken, the register will converge
to the exact value.

\subsection{Circuit Construction}
Let $\ket{\bar{x}_r} = \ket{\bar{x}_{m-1}...\bar{x}_0}$. In order to prepare
the ancilla in the state from~\eqref{eq:phase}, we perform an operation
\begin{align}
\ket{\bar{x}_r}\ket{0}\mapsto
\ket{\bar{x}_r}\left(\sqrt{1-r}\ket{0}+\sqrt{r}\ket{1}\right).
\end{align}
We'll call this phase loading. An example circuit for quantum rounding using
this procedure is described in
Figure~\ref{fig:qr} using the
controlled rotation loading method of~\cite{chakrabarti2020quantum}. Here the
$\ket{\bar{x}_r}$ register undergoes a square root and then arcsin operation.
Controlled rotations are then applied bitwise between this result and another
ancilla.
Unfortunately this method may be resource expensive as it requires computing a
square root and arcsin of a quantum register.

\begin{figure}
\begin{align*}
\Qcircuit @C=0.20em @R=1.25em {
\lstick{\ket{\bar{x}}_n}     &  {/} \qw & \qw      & \qw & \qw & \qw &
\gate{\mbox{ADD}\,\epsilon_{RD}} &
{/} \qw & \qw & \rstick{\ket{x}_n} \\
\lstick{\ket{\bar{x}_{m-1}}} & \multigate{3}{\sin^{-1}\sqrt{\;}} & \ctrl{4} &
\qw & \qw &
 \qw &
\qw & \qw & \qw \\
\lstick{\ket{\bar{x}_{m-2}}} & \ghost{\sin^{-1}\sqrt{\;}} & \qw      & \ctrl{3} & \qw & \qw &
\qw & \qw & \qw \\
\lstick{\vdots}                       & &          & & \ddots & & & & \vdots \\
\lstick{\ket{\bar{x}_{0}}}   & \ghost{\sin^{-1}\sqrt{\;}} & \qw      & \qw & \qw & \ctrl{1} &
\qw & \qw & \qw \\
\lstick{\ket{0}}  & \qw & \gate{RY(\theta_{m-1})} & \gate{RY
(\theta_{m-2})} & \cds{-1}{\cdots} & \gate{RY(\theta_{0})} & \ctrl{-5} &
\meter
}
\end{align*}
\caption{Circuit for quantum rounding an $n+m$-bit number $\ket{\bar{x}}$ to a
$n$-bit number $\ket{x}_n$. Here $\theta_i=\sin^{-1}\left(2^{-(m-i)}\right)$.
The $m$-bit register becomes garbage that can be uncomputed. The ancilla can
be conditionally reset and reused.
}
\label{fig:qr}
\end{figure}

\begin{figure}[ht]
\begin{align*}
\Qcircuit @C=0.20em @R=1.25em {
\lstick{\ket{\bar{x}}_n}     &  {/} \qw & \qw & \qw & \qw & \qw & \qw & \qw &
\gate{\mbox{ADD}\,\epsilon_{RD}} &
{/} \qw & \qw & \rstick{\ket{x}_n} \\
\lstick{\ket{\bar{x}_r}_m} & {/} \qw & \qw & \qw \qw & \multigate{2}{\mbox{Compare}} &\qw & \qw
 & \qw & \qw & \qw & \qw
\\
\lstick{\ket{0}_m} &  {/} \qw & \qw & \gate{H^{\otimes m}} & \ghost{\mbox{Compare}} &\qw
 & \qw & \qw & \qw &
\qw & \qw \\
\lstick{\ket{0}}  & \qw & \qw & \qw & \ghost{\mbox{Compare}} & \qw  & \qw  & \qw  & \ctrl{-3} &
\meter
}
\end{align*}
\caption{Circuit for quantum rounding an $n+m$-bit number $\ket{\bar{x}}$ to a
$n$-bit number $\ket{x}_n$ using the comparator method.
The $m$-bit register becomes garbage that can be uncomputed. The ancilla can
be conditionally reset and reused.
}
\label{fig:comparator}
\end{figure}
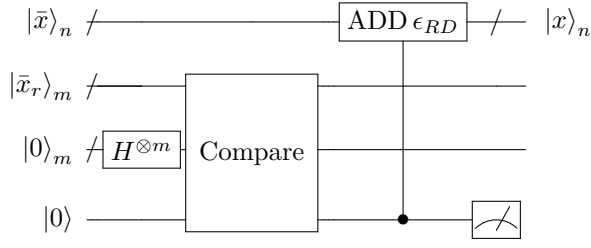

We introduce an alternative method to replace phase loading that is inspired by
quantum counting and that reduces the resources used.~\footnote{Versions of this trick were used for different applications in the
Hamiltonian time evolution simulation
algorithm in~\cite{low2019hamiltonian} and the Coherent Iterative Energy
Estimation algorithm of~\cite{rall2021faster}.}
 We first add
$m$ ancilla qubits in the uniform superpostion.
\begin{align}
\ket{\bar{x}_r}_m\ket{+}_m = \sum_{j=0}^{2^m-1}
\frac{1}{\sqrt{2^m}}\ket{\bar{x}_r}_m\ket{j}_m
\end{align}
We then add an additional qubit controlled on the ancilla and apply a comparator
circuit~\cite{draper2006logarithmic}. The comparator unitary operations perform
\begin{align}
\mbox{Compare}:\ket{a}\ket{b}\ket{0}\mapsto\ket{a}\ket{b}\ket{0\mbox{ if }
a > b \mbox{ and } 1 \mbox{ otherwise}}.
\end{align}
We can think of the $m$ ancilla qubits as a discretization of the interval
$[0, 1]$ into $2^m$ bins. The comparator is then accumulating into a quantum
amplitude an amount $1/\sqrt{2^m}$ for each bin that is less than $r$.
This results in
\begin{align}
\sum_{j=0}^{2^m-1}\frac{1}{\sqrt{2^m}}
\ket{\bar{x}_r}_m\ket{j}_m\ket{j<\bar{x}_r} =
\ket{\bar{x}_r}_m\otimes\left[\sum_{j=0}^{\bar{x}_r-1}\frac{1}{\sqrt{2^m}}\ket{j}_m\ket{1} +
\sum_{j=\bar{x}_r}^{2^m-1}\frac{1}{\sqrt{2^m}}\ket{j}_m\ket{0} \right].
\end{align}
The probabilities of the comparator ancilla outcomes are now
\begin{align}
\mbox{Pr}[\mbox{ancilla is 1}] &= \frac{\bar{x}_r}{2^m} = r \\
\mbox{Pr}[\mbox{ancilla is 0}] &= \frac{2^m-\bar{x}_r}{2^m} = 1-r.
\end{align}
This is precisely what we desire for quantum rounding.
Figure~\ref{fig:comparator} gives an example circuit for this approach.

Note that the mid-circuit measurement of the ancilla qubit does not affect
the operation of any program that uses quantum rounding as a subroutine as long
as the the program's output is eventually classical (rather than a quantum
state output). To see this, we could not re-use ancillas and instead defer all
the mid-circuit
measurements to the end of a quantum computation. The results of the final
ancilla-rounding measurements can then be thrown away while the results of the
computationally relevant measurements are kept.

\subsection{Loading rounded classical numbers}
When the original number $\bar{x}$ is classical, then the procedure can be
simplified. One can use a rotation $RY\left(\sin^{-1}(\sqrt{r})\right)$ on
the largest remainder bit,
but it may be easier to use a purely classical procedure. Instead of
deterministically loading $\bar{x}$, load $\cx$ with probability $r$ and load
$\fx$ with probability $1-r$. Sampling either $\cx$ or $\fx$ can be done
classically. We then gain the advantages of quantum rounding, but use no
additional quantum resources.

\section{Errors in quantum rounding}
\label{sec:qr-errors}
We first consider the case where all gates can be implemented exactly e.g. that
 we are either using
the comparator based method (Fig.~\ref{fig:comparator}) or the rotation method (Fig.~\ref{fig:qr})
with the assumption that we can apply exact $Ry$ gates.
We can use the Chernoff bound to bound the error in the expected value of our
register $\ket{x}$.
\begin{theorem}[Chernoff]
Let $X=\sum_{i=1}^NX_i$ be a sum of independent Bernoulli random variables
$X_i$ that each take the value $1$ with probability $p_i$ and the value $0$
with probability $1-p_i$. Let $\mu = \sum_i^Np_i$ be the mean of $X$. Then
for any choice of $\delta\in(0,1)$ we obtain
\begin{align}
P\left(|X-\mu| \ge \delta\mu\right) \le 2^{-\mu\delta^2/3}.
\end{align}
\end{theorem}

In our setting, $X_i$ is the $i$-th measurement outcome that is sampled from
the rounded state. Then
$p_i=r$ and $\mu=Nr$ and the
expected error after $N$ samples is bounded by
\begin{align}
P\left(\frac{|X-Nr|}{N} \ge \delta r\right) \le 2^{-\mu\delta^2/3}.
\end{align}
Setting $\delta = \sqrt{\frac{3\ln(2/\alpha)}{Nr}}$ gives a failure rate of
$\alpha$ that bounds
\begin{align}
P\left(\frac{|X-Nr|}{N} \ge \sqrt{\frac{3\ln(2/\alpha)}{N}}\right) \le
\alpha,
\end{align}
where we have used the upper bound of $r\in[0,1]$. In practice it may be
reasonable to choose $\alpha = 1/N$ so that a failure
is not expected.

After $N$ samples our empirical mean value for the register is
\begin{align}
x_{QR} = \frac{X}{N}\epsilon_{RD} + \fx.
\end{align}
Let the true value $\bar{x}=\fx + x_R$. Then our error is
\begin{align}
|x_{QR}-\bar{x}| &= |\frac{X}{N}\epsilon_{RD} - x_R| \\
&= |\frac{X}{N} - \frac{x_R}{\epsilon_{RD}}|\epsilon_{RD} \\
&= |\frac{X}{N} - r|\epsilon_{RD} \\
&\le \epsilon_{RD}\sqrt{\frac{3\ln(2/\alpha)}{N}} \qquad \mbox{w/ prob.} \geq
1-\alpha.
\end{align}

\noindent Comparing to the round-nearest method, we reduce the error by a
factor
\begin{align}
F = \frac{1}{2}\sqrt{\frac{3\ln(2/\alpha)}{N}}.
\end{align}
As $N$ increases, the error compared to the round-nearest method decreases
significantly. For $\alpha=1/N$ and $N\approx 500$ then error is reduced by an
order of magnitude. For $\alpha=1/N$ and $N=90$k then error is reduced by two
orders of magnitude.

\subsection{Rotation errors}
If we are using the rotation approach then we should also take into account
errors in rotation gates. Rotation errors may be due to
approximate synthesis in a fault-tolerant setting or physical gate noise in a
NISQ setting. If the errors are unbiased then the Chernoff bounds apply as in
the above section since we still have $\mu = \sum_i^Np_i$.

A bias in the rotations will cause additional error. A rotation error that
results in an amplitude $\sqrt{r}+\epsilon_{rot}$ results in an error on the
probability of $\epsilon_B \le (\sqrt{r}+\epsilon_{rot})^2 \le
2\epsilon_{rot}+\epsilon_{rot}^2$. Thus we obtain
\begin{align}
|x_{QR}-\bar{x}| \le \epsilon_{RD}\left(\sqrt{\frac{3\ln(2/\alpha)}{N}} +
\epsilon_B\right).
\label{eq:error_with_ry_error}
\end{align}

\section{Resource Estimates}
\label{sec:overview-resources}

We now consider the resources used to implement quantum rounding circuits.
In the fault-tolerant setting we assume that resources are dominated by the T gate count and depth, with the CNOT
gate count and depth having a smaller but still non-trivial impact. Therefore we choose to
use appropriate circuits that minimize the T-depth whilst still ensuring that our CNOT gate counts
and depth do not become exponentially large. A more detailed discussion of the focus on minimizing the T-depth 
 can be found in Section~\ref{subsec:ft_resources}.

Table~\ref{tab:first} shows the resources used
for a $10$-bit representation of $\bar{x}$ along
with a $10$-bit remainder, i.e. when $n=m=10$. The method used is the
comparator circuit from Figure~\ref{fig:comparator}.
These tables also show the leading order dependence for generic $n$ and $m$.
Additional resource estimates along with a more detailed explanation of their
computations are given in~\autoref{sec:resources}.

\begin{table}[th]

\centering
\caption{Resources for quantum rounding in a fault-tolerant setting. The
rounding circuit used is in~Fig.\ref{fig:comparator}.}
\label{tab:first}

\begin{tabular}{| l | c |c |}
\hline
Resource & $n=m=10$ & Leading order terms\\ \hline
\hline
Additional qubits & 12 & $m$\\
\hline
Uncomputed ancillas & 34 & $\max(4m, 3n)$\\
\hline
T-count & 588 & $24m+30n$\\
\hline
T-depth & 32 & $2\log_2(m)+2\log_2(n)$\\
\hline
CNOT count & 1509 & $62m+82n$\\
\hline
CNOT depth & 323 & $20\log_2(m)+20\log_2(n)$\\
\hline
\end{tabular}

\end{table}

\section{Applications and Benchmarks}
\label{sec:benchmarks}

For a single rounding step, the error can be exponentially decreased by increasing
the rounded register size $n$. This might appear to be preferable to
the overhead of quantum rounding. However, increasing precision means
that the rest of the computation must proceed with a larger value of $n$
after that rounded step. If the rest of the computation is long and/or involved
then benefits can be found from quantum rounding.

\subsection{Fixed point multiplication benchmarks}

As an example, we calculate the resources for a fixed-point
multiplication in a fault tolerant setting, where several steps of rounding are
often used. For a number of samples $N>1$ quantum rounding will add extra
accuracy per rounding step, allowing us
to reduce the precision to achieve a target overall accuracy. In our
benchmarks we choose $n=m$ for simplicity and choose the target $\epsilon =
n/2^{n-p}$ as this is the baseline accuracy from the Haner et al.
method from Appendix A in~\cite{haner2018optimizing}.~\footnote{This method can
be described as
the binary version of the usual schoolbook multiplication i.e. a series of additions of one number register to a
result register controlled on the other number register. However in the original paper, the authors use the addition circuit in \cite{takahashi2009quantum}
when computing the resources for this method of multiplication. We instead modify it slightly by using the addition circuit in 
\cite{draper2006logarithmic} which is optimized for T-depth (at the cost of an increase in T-count). Going forward
we will refer to this modified version as the Haner et al. method.}

\begin{figure}[h!t]
    \centering
    \begin{subfigure}[b]{0.48\textwidth}
        \centering
        \includegraphics[width=\textwidth]{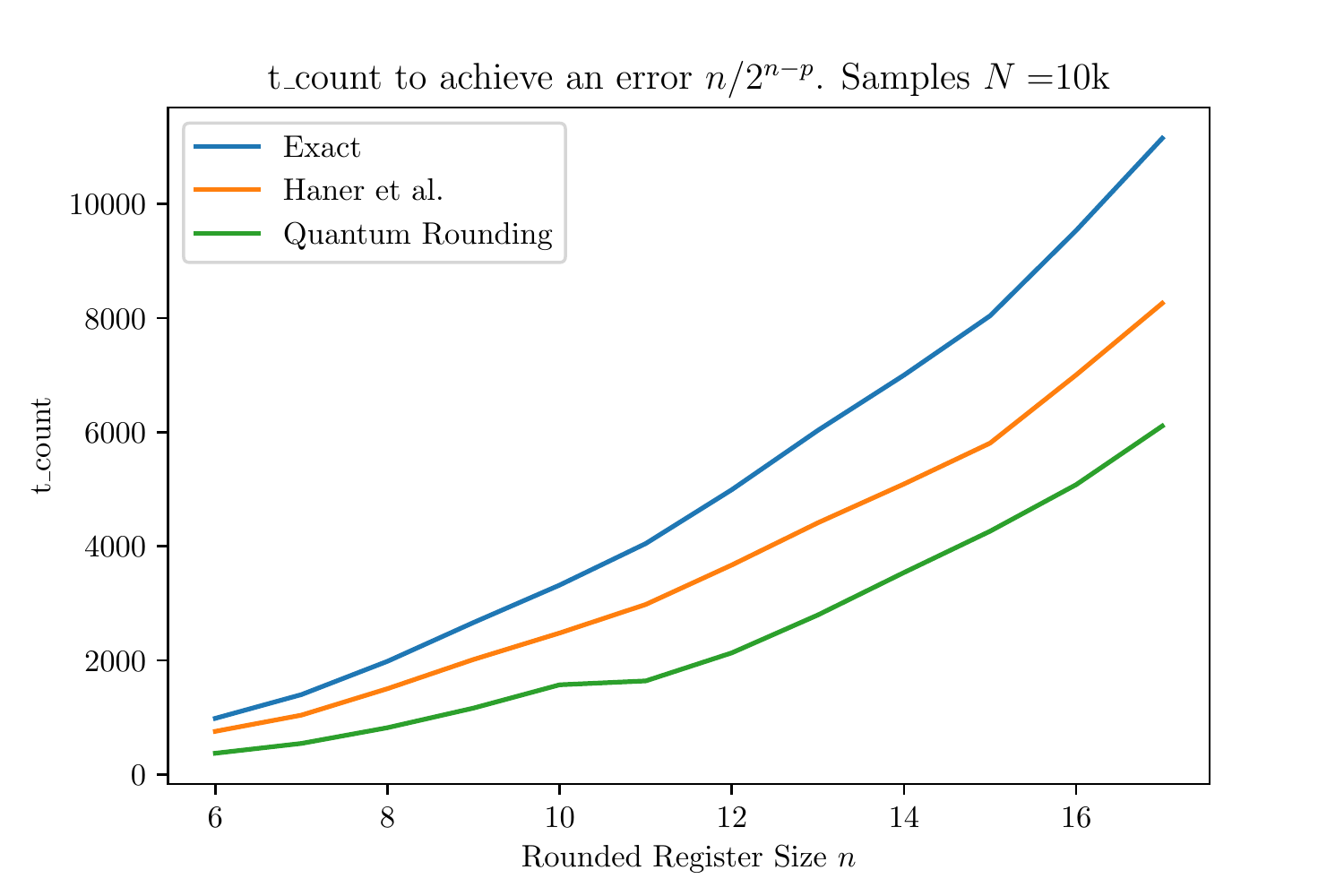}
        \caption{}
        \label{fig:size_tcount}
    \end{subfigure}
    \hfill
    \begin{subfigure}[b]{0.48\textwidth}
        \centering
        \includegraphics[width=\textwidth]{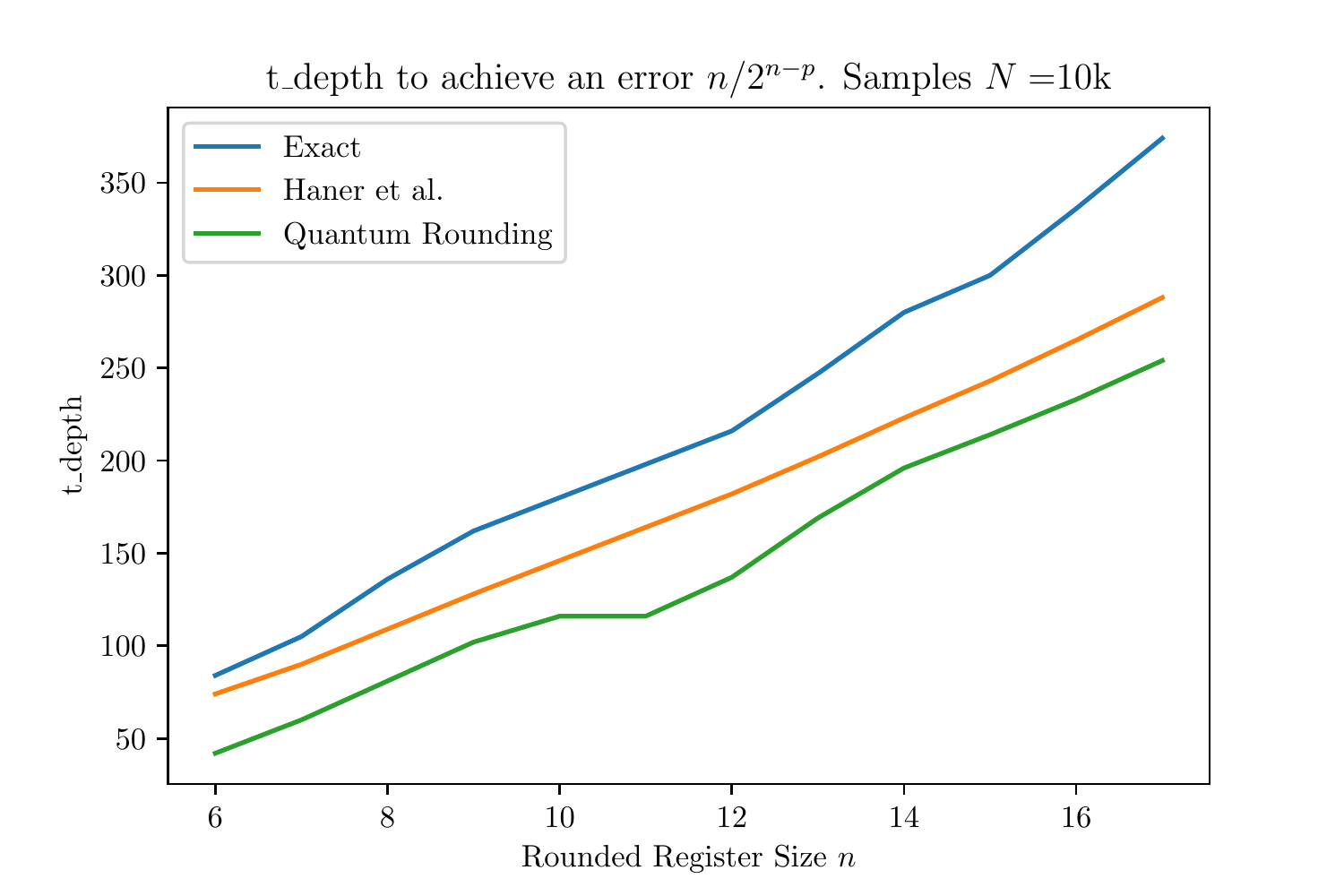}
        \caption{}
        \label{fig:size_tdepth}
    \end{subfigure}
    \hfill
     \begin{subfigure}[b]{0.48\textwidth}
        \centering
        \includegraphics[width=\textwidth]{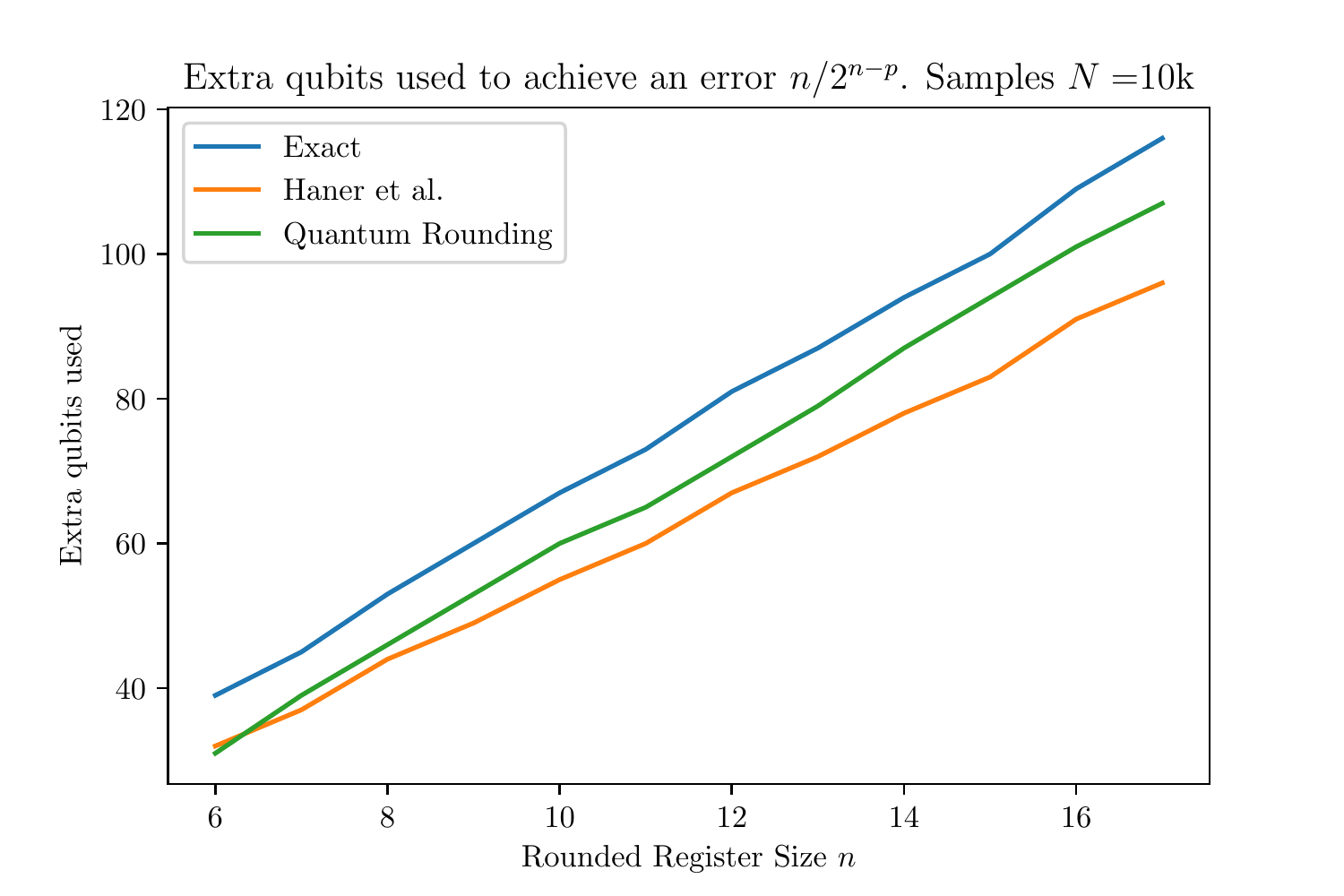}
        \caption{}
        \label{fig:size_qubits}
    \end{subfigure}
     \hfill
     \begin{subfigure}[b]{0.48\textwidth}
        \centering
        \includegraphics[width=\textwidth]{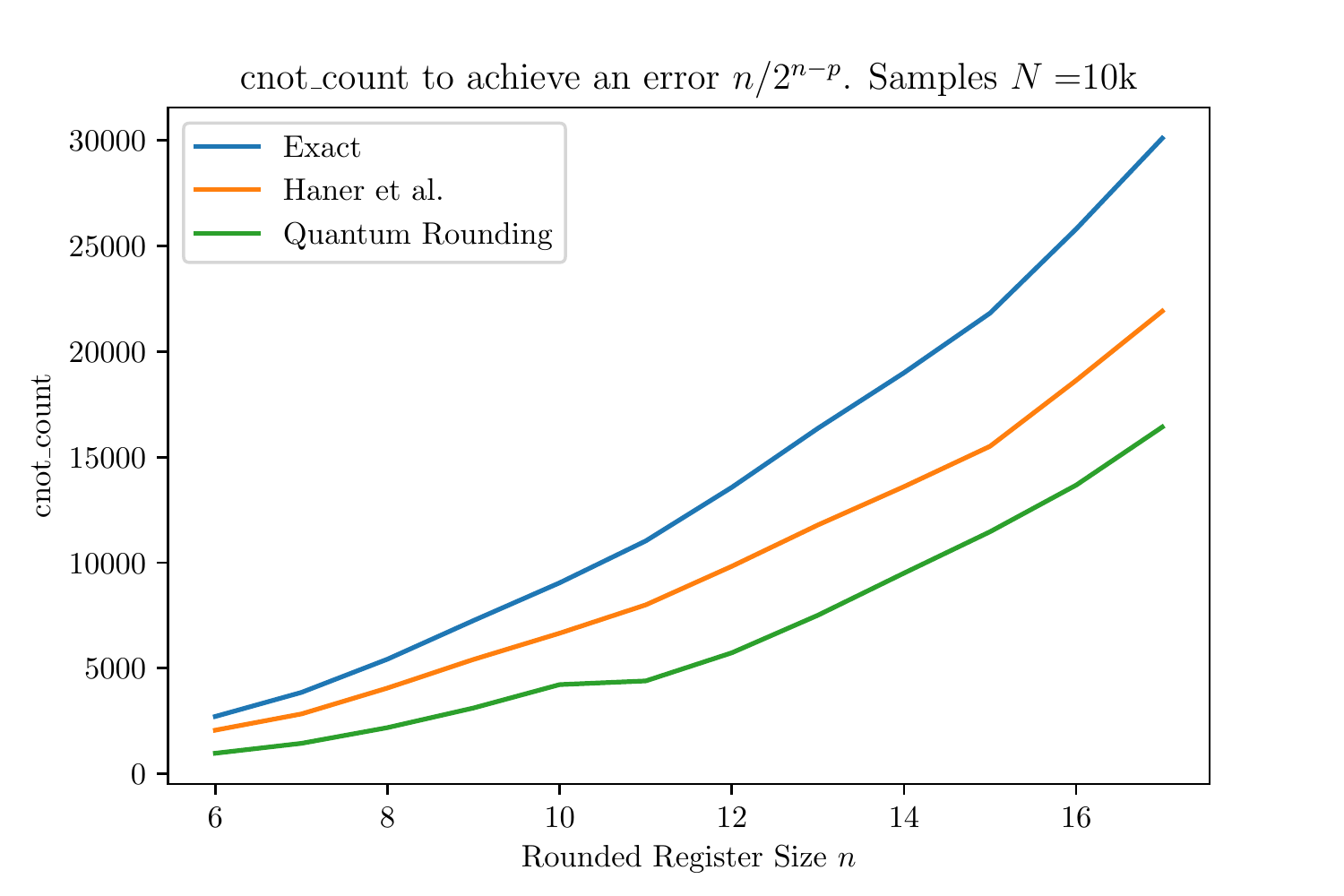}
        \caption{}
        \label{fig:size_cnotcount}
    \end{subfigure}
    \hfill
     \begin{subfigure}[b]{0.48\textwidth}
        \centering
        \includegraphics[width=\textwidth]{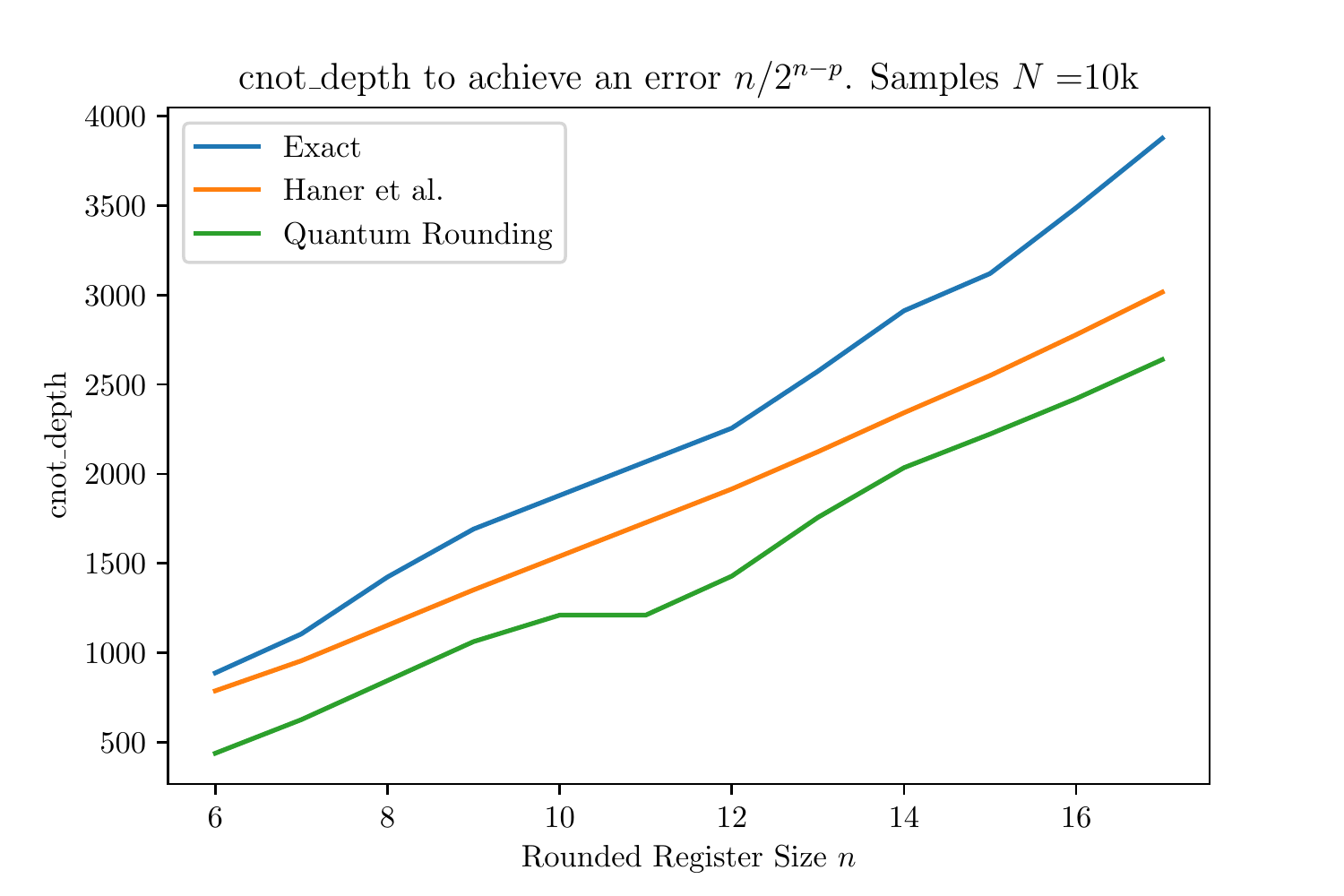}
        \caption{}
        \label{fig:size_cnotdepth}
    \end{subfigure}

	\caption{Comparison of different resource metrics for fixed point
	multiplication at different register sizes $n$. We assume $N=10$k samples
	for the quantum rounding method. The resources estimated are for a
	fault-tolerant implementation of quantum rounding using the comparator
	method from Fig.~\ref{fig:comparator}. Descriptions of the three methods
	are given in the text. We see reductions in the number of
	gates and their depth for the quantum rounding approach. The number
	of qubits used is slightly worse for quantum rounding.
	}
	\label{fig:sizes}
\end{figure}

\begin{figure}
    \centering
    \begin{subfigure}[b]{0.48\textwidth}
        \centering
        \includegraphics[width=\textwidth]{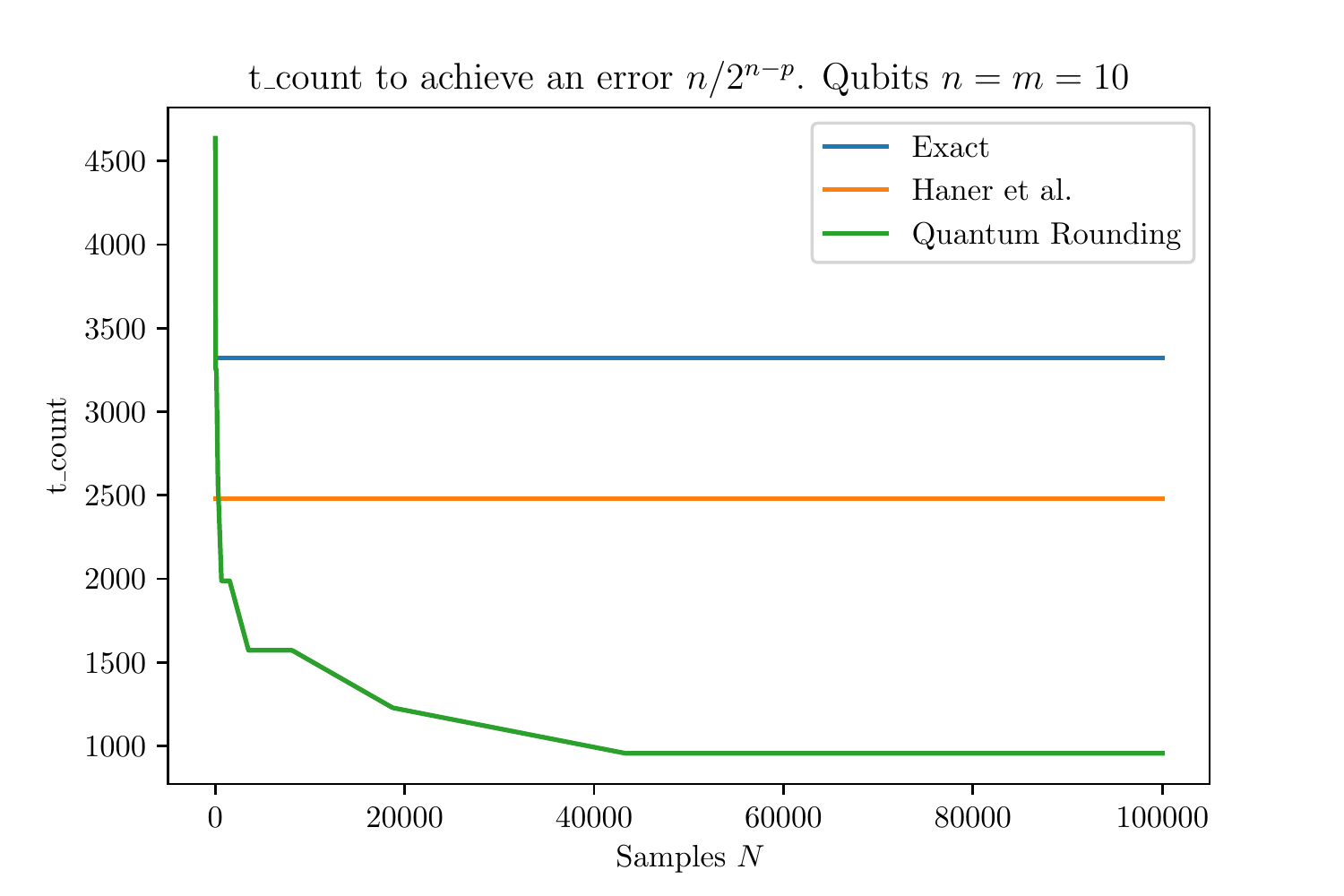}
        \caption{}
        \label{fig:samples_tcount}
    \end{subfigure}
    \hfill
    \begin{subfigure}[b]{0.48\textwidth}
        \centering
        \includegraphics[width=\textwidth]{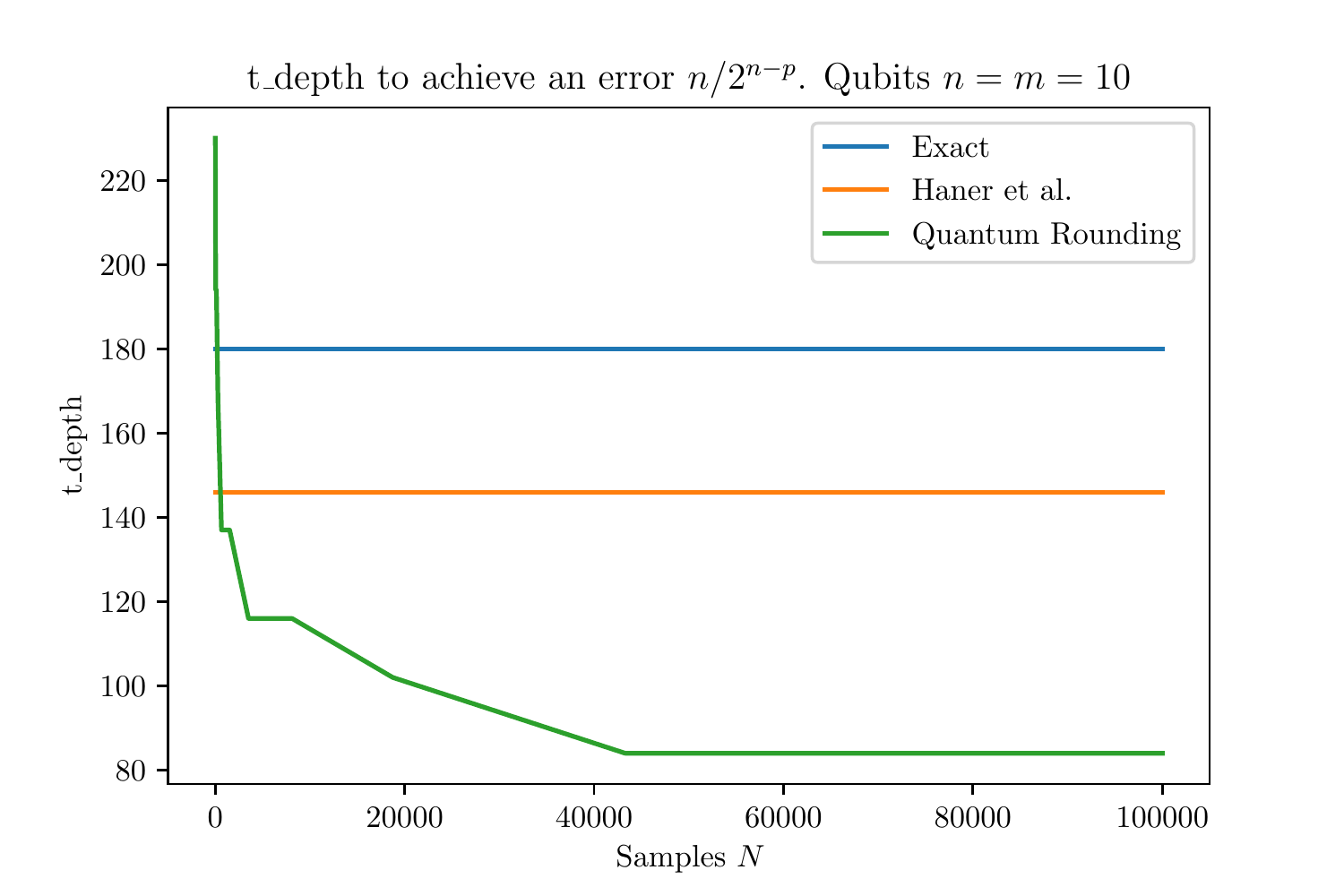}
        \caption{}
        \label{fig:samples_tdepth}
    \end{subfigure}
    \hfill
     \begin{subfigure}[b]{0.48\textwidth}
        \centering
        \includegraphics[width=\textwidth]{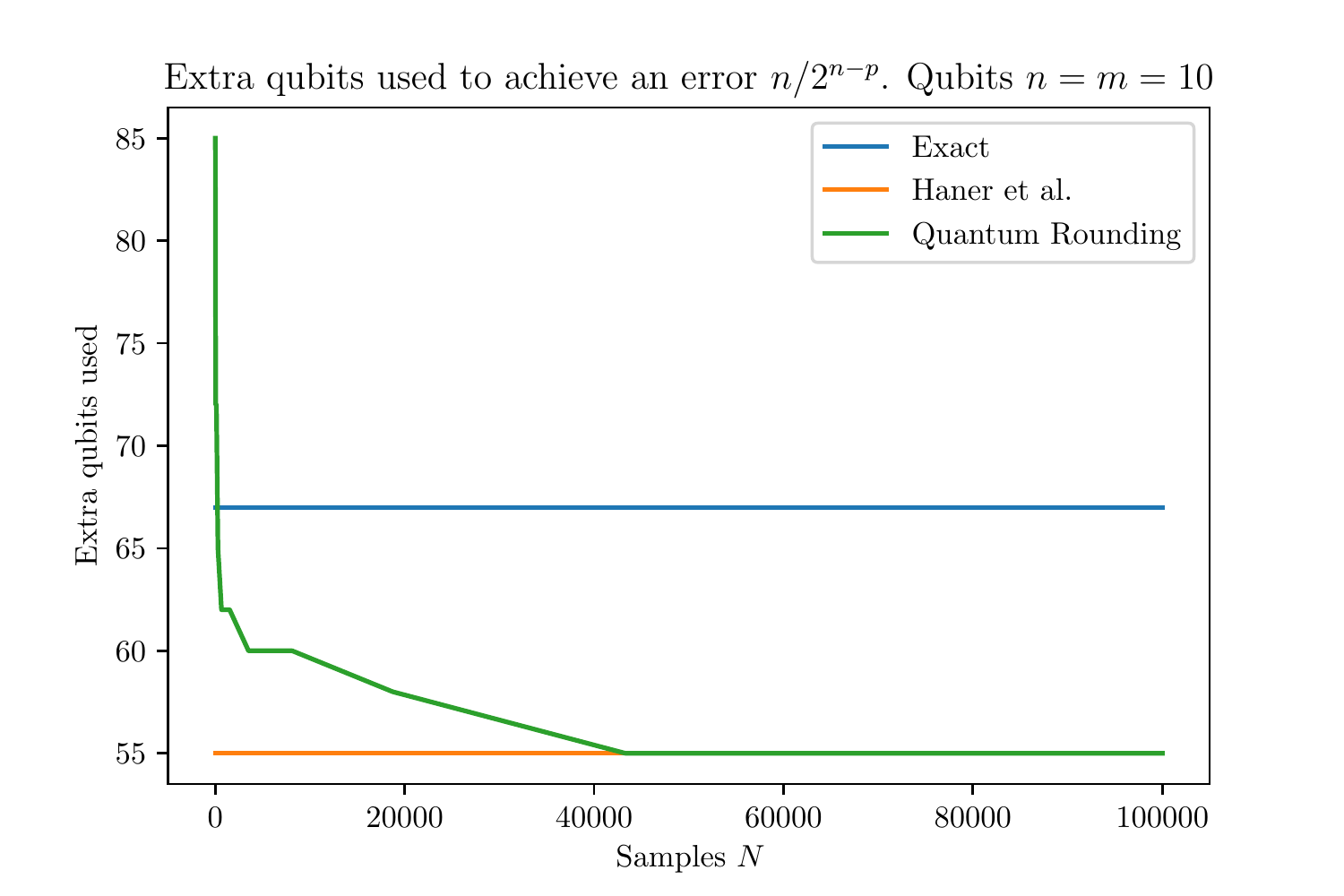}
        \caption{}
        \label{fig:samples_qubits}
    \end{subfigure}
     \hfill
     \begin{subfigure}[b]{0.48\textwidth}
        \centering
        \includegraphics[width=\textwidth]{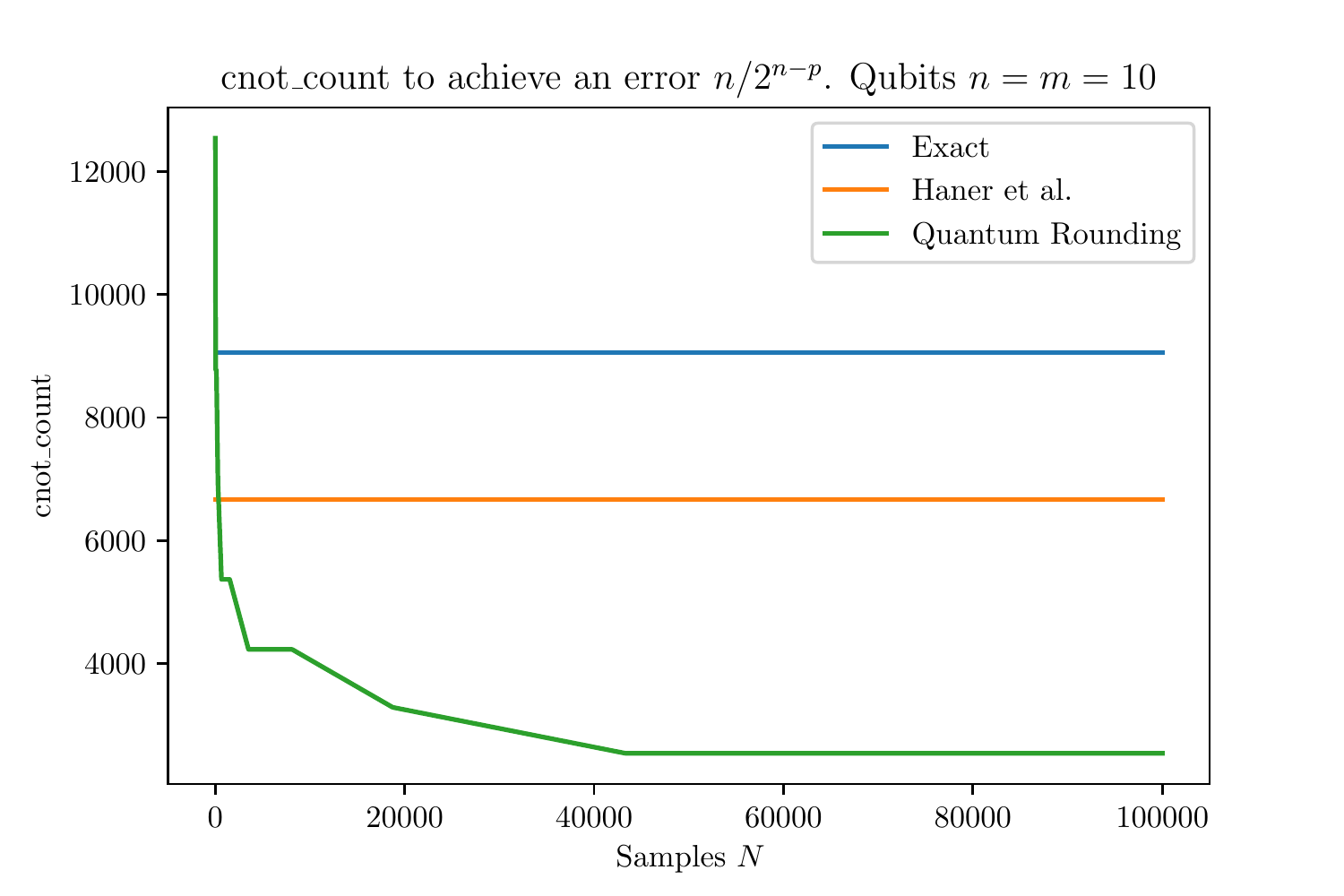}
        \caption{}
        \label{fig:samples_cnotcount}
    \end{subfigure}
    \hfill
     \begin{subfigure}[b]{0.48\textwidth}
        \centering
        \includegraphics[width=\textwidth]{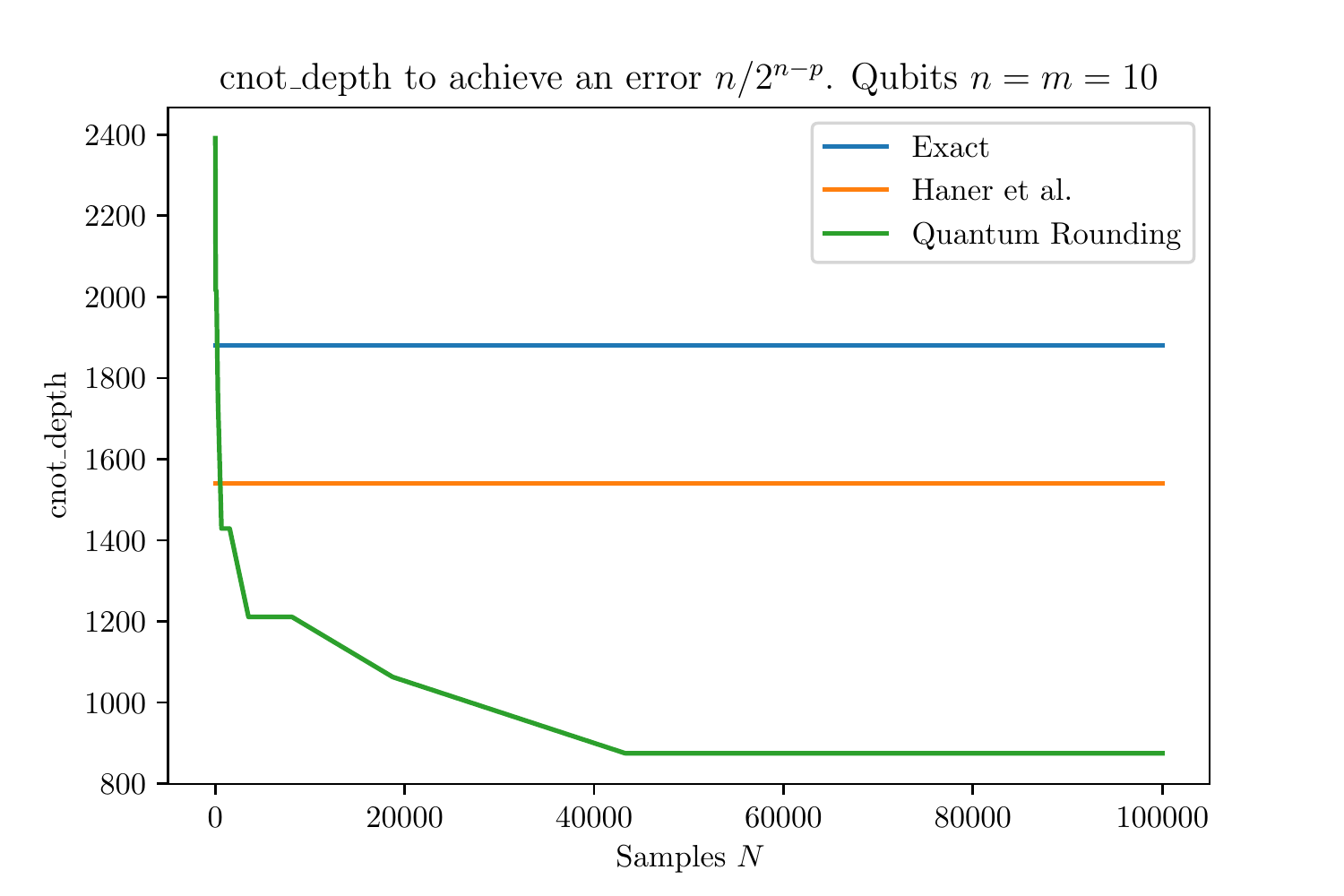}
        \caption{}
        \label{fig:samples_cnotdepth}
    \end{subfigure}

	\caption{Comparison of different resource metrics for fixed point
	multiplication with a different number of samples $N$. We assume
	$n=10$-bit quantum registers. The resources estimated are for a
	fault-tolerant implementation of quantum rounding using the comparator
	method from Fig.~\ref{fig:comparator}. Descriptions of the three methods
	are given in the text. We see reductions in the number of
	gates and their depth for the quantum rounding approach. The number
	of qubits used is worse for quantum rounding when $N<40$k. For larger
	$N$ the qubit number becomes comparable. We see that the resource
	improvement from quantum rounding approaches a floor for $N$ of
	approximately $50$k samples in this case. For very small $N$ (e.g.
	$N < 50$) quantum rounding is the worst method, but rapidly improves as the
	number of samples grows.}
	\label{fig:samples}
\end{figure}

In Figure~\ref{fig:sizes} and Figure~\ref{fig:samples} we compare using three
different techniques on $n$-bit fixed
point multiplication. The Exact method is
exact multiplication to a $2n$-bit precision that is then rounded-down
to $n$-bit precision. The Haner et al.~\cite{haner2018optimizing} method uses
fewer qubits per addition and does no rounding since it assumes the worst case
error for each addition. A detailed explanation of the implementations of these
methods is given in \autoref{sec:fpm}. Our quantum rounding method
does exact
multiplication to $2\tilde{n}$-bit precision and then quantum rounds
down to $\tilde{n}$-bit precision. Here $\tilde{n}$ is the smallest register
size that matches the error target $\epsilon= n/2^{n-p}$ of the other
approaches. Generally $\tilde{n}\le n$ with the difference growing as the
number of samples $N$ increases. We calculate resources using the
fault-tolerant resource counts for quantum rounding.

Figure~\ref{fig:sizes} compares the different methods for a fixed number of
samples $N=10$k across different register sizes $n$. For the gate resources of
CNOT and T gate counts and depths, quantum rounding reduces resources by about a
factor of 2. However, quantum rounding uses slightly more qubits than Haner
et al. (but fewer than the Exact method). This is shown in
Figure~\ref{fig:size_qubits}.

Figure~\ref{fig:samples} fixes the register size to $10$ qubits and varies the
number of samples. We see that across the gate metrics more samples rapidly
reduce the resources used. This advantage plateaus at about a 3X advantage
over the Haner et al. method. We see that the number of qubits used starts
worse than the Haner et al. method but becomes comparable for large numbers of
samples.

These benchmarks give just one example using fixed-point multiplication. For
other subroutines and larger algorithms the optimal tradeoff between qubits,
additions, and rounding accuracy will likely vary.

\subsection{Other applications}

The idea behind quantum rounding can also be applied to
floating point representations where we wish to better round between machine
representable numbers. Here one could apply quantum rounding methods to the
mantissa.

One could also consider approximate versions of quantum rounding. In
Appendix~\ref{sec:quantum-semi} we consider a version of quantum rounding
-~\emph{quantum semi-rounding} - where
only the most significant bit of the remainder is used. This method does not
appear to result in a saving over the comparator implementation of quantum
rounding due to the resources used just to estimate the most significant
bit. There may still be a better method for this approach.

Quantum rounding methods are likely to be particularly relevant in
applications of quantum computers to problems that have classical
oracles such as in optimization and finance. In those cases the quantum
implementation of the arithmetic in the classical oracle can absorb significant
resource costs, e.g. in the Quantum Approximate Optimization
Algorithm~\cite{farhi2014quantum} applied
to classical objective functions.

\section{Acknowledgements}
We would like to thank Thomas H{\"a}ner and Dimitri Maslov for
references and suggestions on quantum arithmetic constructions. We thank
Patrick Rall and Nikitas Stamatopoulos for helpful discussion and review.

\bibliographystyle{plain}
\bibliography{references}

\appendix

\section{Quantum semi-rounding}
\label{sec:quantum-semi}

\begin{figure}
	\begin{align*}
	\Qcircuit @C=0em @R=0.9em {
	\lstick{\ket{\bar{x}}_n} & {/} \qw 				     & \qw        & \qw &
					\qw      				     & \qw        & \qw &
					\qw &  {/} \qw & \qw &
						\qw       & \qw       & \qw      & \qw &
						\qw       & \qw       & \qw      &
				\qw  & \gate{\mbox{ADD}\,\epsilon_{RD}} & \qw &
					\rstick{\ket{x}_n} \\
	\lstick{\ket{\bar{x}_{m-1}}} & \ctrl{17}  & \gate{X} & \ctrl{1}  &
					       \qw  	 & \ctrl{1}    & \qw       &
					       \qw         & \qw          & \qw       &
						\qw       & \qw       & \qw      & \qw &
						\qw       & \qw       & \qw      &
						\qw       & \qw       & \qw      \\
	\lstick{\ket{\bar{x}_{m-2}}} & \qw         & \qw          & \ctrl{6}  &
					       \gate{X} & \ctrl{7}    & \qw        &
					       \qw         & \qw          & \qw &
						\qw       & \qw       & \qw      & \qw &
						\qw       & \qw       & \qw      &
						\qw       & \qw       & \qw      \\
	\lstick{\ket{\bar{x}_{m-3}}} & \qw         & \qw          & \qw        &
					       \qw         & \qw          & \ctrl{6}  &
					       \gate{X} & \ctrl{6}    & \qw        &
                                                            \qw         & \qw          & \qw        & \qw &
					       \qw       & \qw       & \qw      &
				    	       \qw       & \qw       & \qw      \\
	\lstick{\ket{\bar{x}_{m-4}}} & \qw         & \qw          & \qw        &
					       \qw         & \qw          & \qw	     &
					       \qw         & \qw          & \ctrl{7}  &
					       \gate{X}  & \qw          & \qw       & \qw &
					       \qw       & \qw       & \qw      &
				    	       \qw       & \qw       & \qw      \\
	\lstick{\vdots} & & 	          &             &
			      &	          &             &
			      &            &             &
			      & \ddots &             &  &
			      &             &            &
			      &             & \vdots \\
	\lstick{\ket{\bar{x}_{1}}}   &  \qw         & \qw          & \qw       &
					     \qw        & \qw          & \qw       &
					     \qw        & \qw          & \qw       &
					     \qw        & \qw          & \qw       & \qw &
					     \qw  &  \ctrl{9}   & \qw       &
 					    \qw        & \qw          & \qw       \\
	\lstick{\ket{\bar{x}_{0}}}   &  \qw         & \qw          & \qw       &
					     \qw        & \qw          & \qw       &
					     \qw        & \qw          & \qw       &
					     \qw        & \qw          & \qw       & \qw &
					     \qw        &  \qw         & \ctrl{9}   &
 					    \qw        & \qw          & \qw       \\
	\lstick{\ket{a_{m-2}}}    & \qw       & \qw       & \targ 	   &
					\ctrl{10} & \qw       & \qw 	   &
					\qw 	  & \qw       & \qw       &
					\qw 	  & \qw      & \qw   & \qw &
					\qw       & \qw       & \qw      &
					\qw       & \qw       & \qw      \\
	\lstick{\ket{a_{m-3,0}}} & \qw 	  & \qw       & \qw       &
					\qw 	  & \targ     & \ctrl{1} &
					\qw 	  & \ctrl{2} & \qw      &
					\qw 	  & \qw      & \qw  & \qw &
					\qw       & \qw       & \qw      &
					\qw       & \qw       & \qw      \\
	\lstick{\ket{a_{m-3,1}}} & \qw 	  & \qw       & \qw	  &
					\qw 	  & \qw       & \targ 	  &
					\ctrl{8} & \qw       & \qw      &
					\qw 	  & \qw      & \qw  & \qw &
					\qw       & \qw       & \qw      &
					\qw       & \qw       & \qw      \\
	\lstick{\ket{a_{m-4,0}}} & \qw 	  & \qw       & \qw	  &
					\qw 	  & \qw       & \qw 	  &
					\qw	  & \targ     & \ctrl{1} &
					\qw       & \qw       & \qw & \qw &
					\qw       & \qw       & \qw      &
					\qw       & \qw       & \qw      \\
	\lstick{\ket{a_{m-4,1}}} & \qw 	  & \qw       & \qw	  &
					\qw 	  & \qw       & \qw 	  &
					\qw       & \qw       & \targ    &
					\ctrl{6} & \qw       & \qw  & \qw &
					\qw       & \qw       & \qw      &
					\qw       & \qw       & \qw      \\
	\lstick{\vdots} & & 	          &             &
			      &	          &             &
			      &            &             &
			      & \ddots &             &  &
			      &             &            &
			      &             & \vdots \\
	\lstick{\ket{a_{1,0}}}    & \qw 	  & \qw       & \qw	  &
					\qw 	  & \qw       & \qw 	  &
					\qw       & \qw       & \qw      &
					\qw       & \qw       & \qw      & \qw &
					\qw       & \ctrl{2} & \qw      &
					\qw       & \qw       & \qw      \\
	\lstick{\ket{a_{1,1}}}    & \qw 	  & \qw       & \qw	  &
					\qw 	  & \qw       & \qw 	  &
					\qw       & \qw       & \qw      &
					\qw       & \qw       & \qw  & \qw &
					\qw       & \qw       & \qw      &
					\qw       & \qw       & \qw      \\
	\lstick{\ket{a_{0,0}}}    & \qw 	  & \qw       & \qw	  &
					\qw 	  & \qw       & \qw 	  &
					\qw       & \qw       & \qw      &
					\qw       & \qw       & \qw  & \qw &
					\qw       & \targ     & \ctrl{1} &
					\qw       & \qw       & \qw      \\
	\lstick{\ket{a_{0,1}}}    & \qw 	  & \qw       & \qw	  &
					\qw 	  & \qw       & \qw 	  &
					\qw       & \qw       & \qw      &
					\qw       & \qw       & \qw  & \qw &
					\qw       & \qw       & \targ    &
					\ctrl{1} & \qw       & \qw\\
	\lstick{\ket{0}} & \gate{RY(\theta_{m-1})} & \qw & \qw &
		          	       \gate{RY(\theta_{m-2})} & \qw & \qw &
			       \gate{RY(\theta_{m-3})} & \qw & \qw &
			       \gate{RY(\theta_{m-4})} & \qw  & \qw & \cds{-1}{\cdots}  &
			       \qw     & \qw & \qw &
			       \gate{RY(\theta_{0})}     & \ctrl{-18} & \meter
	}
	\end{align*}
\caption{
Circuit for quantum semi-rounding an $n+m$-bit number $\ket{\bar{x}}$
to a $n$-bit number $\ket{x}_n$. Here $\theta_i=\sin^{-1}\left(2^{-(m-i)
}\right)$. The $m$-bit register becomes garbage that can be uncomputed. The
ancilla can be conditionally reset and reused.
The Boolean logic of the Toffoli gates means that each rotation $Ry(\theta_i)$
is only applied if
$\bar{x}_i$ is the most significant $1$-bit in the $m$-bit register.
}
\label{fig:qsr}

\end{figure}
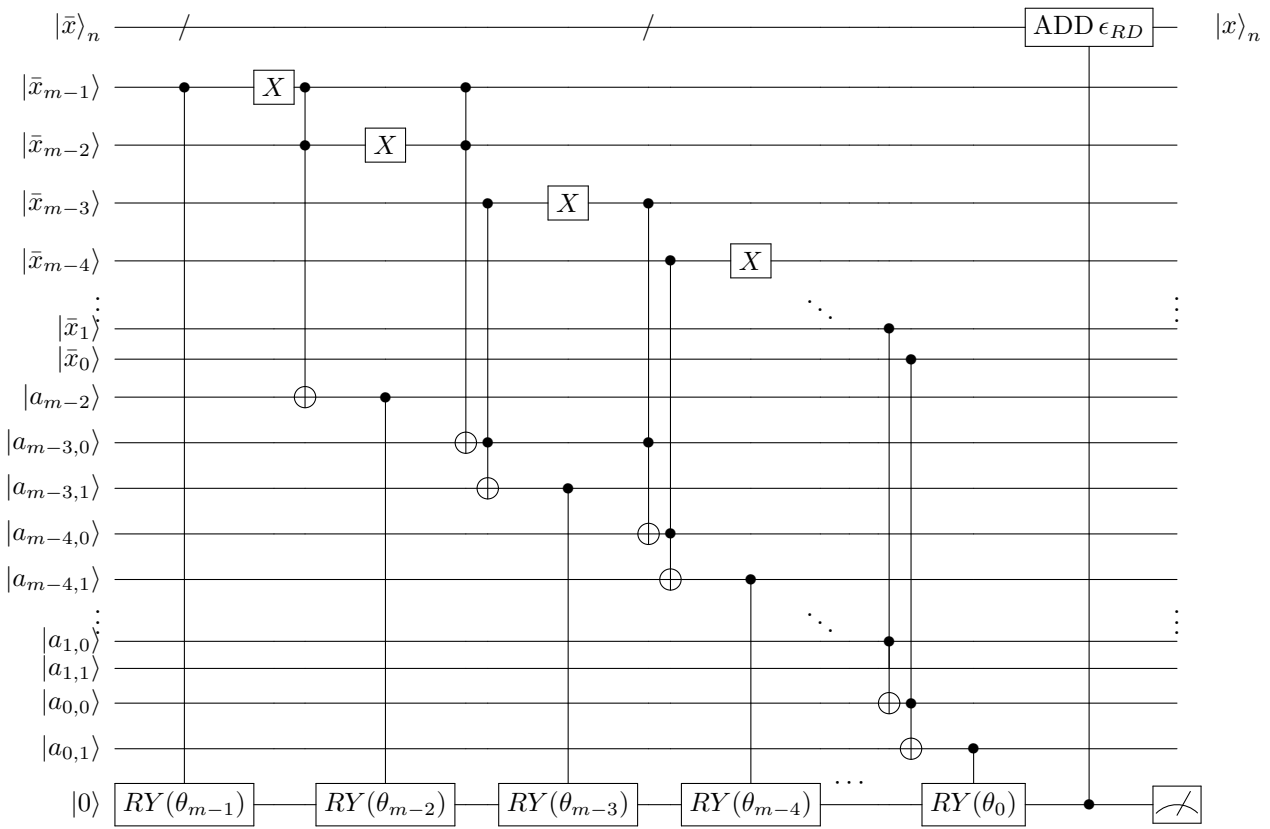

In this section we introduce \emph{quantum semi-rounding} to gain some of the
benefits of quantum rounding but with potentially less quantum resources.
In this variant, we only round the most significant bit of the remainder
$\bar{x}_{m-1}...\bar{x}_0$. An example circuit for this is illustrated in
Fig.~\ref{fig:qsr}, however other circuits may be used. In this method, we have
no error when the remainder has a Hamming weight of one and have less error on
other remainders. Quantum semi-rounding does not currently improve on the
quantum rounding implementation with the comparator, but if there were to be an
improved quantum circuit to identify the most significant bit in a register
then it potentially could.

While quantum rounding improves the worst case error, quantum semi-rounding
does not. The worst case error is the same as round-nearest where we consider
only the most significant bit $\bar{x}_{m-1}$. However, the average case will
be better. If we assume that we are rounding across uniformly distributed
remainders then the average case error from rounding down is
\begin{align}
\hat{\epsilon}_{RD} &= \frac{1}{2^m}\sum_{r}r =
\frac{1}{2^m}\sum_{j=0}^{2^m-1}\left(j\frac{\epsilon_{RD
}}{2^m}\right) \\
&= \frac{\epsilon_{RD}}{2^{m+1}}\left(2^m-1\right).
\end{align}

For quantum semi-rounding we index the error
contribution of each remainder based on the number $k$ of left padding zeros.
The uniform average case error for quantum semi-rounding (for $m\ge
 2$) is given by
\begin{align}
\hat{\epsilon}_{QSR} &= \frac{1}{2^m}\sum_{k=1}^{m-1}\sum_{j=0}^{2^{m-k}-1
}\left(j\frac{\epsilon_{RD}}{2^m}\right) \label{eq:qsr_first_step}\\
&= \frac{\epsilon_{RD}}{3\cdot 2^{2m+1}}\left(4^m-3\cdot 2^m +2\right).
\end{align}
Their ratio is then
\begin{align}
\frac{\hat{\epsilon}_{QSR}}{\hat{\epsilon}_{RD}} = \frac{1}{3 \cdot 2^m}
\left(2^m-2\right).
\end{align}
For large $m$ this approaches an average case error improvement of
3X compared to conventional rounding methods. For small $m$ we do better, with
the $m=2$ and $m=3$ cases resulting in improvements of 6X and 4X respectively.

These values are the average case errors that quantum semi-rounding
converges to in the limit of many samples. This contrasts with the quantum
rounding case where the limit of large samples converges to zero error.
Similarly we expect to converge with large $N$ so that
\begin{align}
\hat{\epsilon}_{QSR} \le \mathcal{O}
\left(\frac{\hat{\epsilon}_{RD}\left(2^m-2\right)}
{3 \cdot 2^m}+\frac{1}{ \sqrt{N}}\right),
\end{align}
for a fixed probability of failure and a number of samples $N$.

\subsection{Quantum semi-rounding with $l$ most significant bits of the
remainder}
\label{sec:qsr_l}
We can also extend the quantum-semi rounding method to determine the rotation angle based on
the $l$ most significant bits of the remainder instead of just the first. This can be done by performing the appropriate Boolean logic before
each controlled rotation. The circuit to implement this is an extension of the one showed in Figure \ref{fig:qsr}
(which depicts the case when $l=1$) where we add the appropriate Toffoli gates and ancilla qubits to compute
the aforementioned Boolean logic before each controlled rotation gate. The
Toffoli-count and depth for this method increases exponentially with $l$
and hence this extension may quickly become inefficient compared to the full
quantum rounding method, e.g. for values $l>3$. However, this method leads to an enhancement in the 
performance in the average case error. The goal is to find the first non-zero bit as we scan the remainder string from
$\ket{\bar{x}_{m-1}}$ to $\ket{\bar{x}_0}$. To compute
this, we start at Eq.~\eqref{eq:qsr_first_step}
but instead of starting the second sum going from $j=0$ to $j=2^{m-k}-1$, it
goes from $j=0$ to $j=2^{m-k-l+1}-1$ which gives us
\begin{align}
\hat{\epsilon}_{QSR_l} &= \frac{1}{2^m}\sum_{k=1}^{m-1}\sum_{j=0}^{2^{m-k-l+1}-1
}\left(j\frac{\epsilon_{RD}}{2^m}\right) \label{eq:qsr_first_step}\\
&= \frac{\epsilon_{RD}}{3\cdot 2^{2m+l}}\left(2\cdot 2^{2m-l}-3\cdot 2^m - 8 \cdot 2^{-l}+6\right).
\end{align}

Following the same procedure as in the main text, we expect to converge to this error so that
\begin{align}
\hat{\epsilon}_{QSR_l} \le \mathcal{O}\left(\frac{\hat{\epsilon}_{RD}\left(2^{1-l}(2^{2m-l}
+3\cdot 2^m+2^{2m-l}+6)\right)}{3\cdot (4^m-2^m)}+\frac{1}{\sqrt{N}}\right),
\end{align}
for a fixed probability of failure and a number of samples $N$.

\section{Resource Estimation Framework}
\label{sec:resources}
In this section we explicitly compute the resources that may be used to run
the quantum rounding and quantum semi-rounding algorithms in both a
fault-tolerant and NISQ setting.

\subsection{Computing resources in a Fault-Tolerant setting}
\label{subsec:ft_resources}
In a fault-tolerant setting, we assume that resources are dominated by 
the T gate count and depth, with the CNOT
gate count and depth having a smaller but still non-trivial impact. Therefore we choose to
use appropriate circuits that minimize the T-depth whilst still ensuring that our CNOT gate counts
and depth does not become exponentially large. This last constraint comes from the fact that any
Boolean function of $n$ qubits can be constructed with $\mathcal{O}(\log(n))$ T-depth but $\mathcal{O}(2^n)$
CNOT gates and ancilla qubits, by decomposing the function using the Positive Polarity Reed-Muller expansion \cite{reed1954class,muller1954application}
where the Boolean products of $n$ variables can be done with T-depth $\log(n)$. However the quantum implementation
of the XORs in the expansion requires a CNOT depth of $\mathcal{O}(2^n)$. In addition this construction also requires $\mathcal{O}(2^n)$
ancilla qubits. Therefore we explicitly avoid circuits that are constructed in this manner.

We use the following calculations:
\begin{itemize}
	\item A single Toffoli gate has a T-count of 4, a T-depth of 1,
	a CNOT count and depth of 10 and uses
	 2 ancilla qubits which are uncomputed (from Figure 1 in
	 \cite{jones2013low}).
	\item The circuit we will use for the $Ry$ rotations is the repeat-until-success one described in \cite{bocharov2015efficient,kliuchnikov2013fast}. When taking into account
	the probability of failure, this results in an average T-count and
	T-depth of $\lceil 1.149\log_2(1/\epsilon) +9.2\rceil$ from
	\cite{ross2016optimal} where $\epsilon$ is the precision of the rotation angle.
	\item The controlled-$Ry$ rotation with the least
	T-depth can be found in \cite{amy2013meet,choi2018efficient}, which gives
	a T-count of $3 \times \lceil 1.149\log_2(1/\epsilon) +9.2\rceil$, a
	T-depth of
	$\lceil 1.149\log_2(1/\epsilon) +9.2 \rceil$, a CNOT count and depth of 4
	 and an additional uncomputed ancilla qubit for a given error $\epsilon$.
	\item The circuit to compare two registers has a Toffoli count of $6n -
	2\lceil\log_2(n)\rceil - 4$, a Toffoli depth of $2\lceil\log_2(n)\rceil + 5$
	with an additional $2n-2$ CNOT count and $2$ CNOT depth (from Table 1 in \cite{draper2006logarithmic}).
\end{itemize}

\subsection{Resources in a NISQ setting}
In a NISQ setting, we treat $Ry$ as a native gate.\footnote{In practice
hardware may make available some other parameterized single-qubit rotation, but
transforming between them may require only a small basis change.}
We focus on quantifying the 2-qubit and single-qubit gate counts and depths.
Therefore, we choose the more NISQ suitable circuits for certain subroutines
that reduce (e.g. minimize) the 2-qubit gate depth.
\begin{itemize}
	\item We treat controlled-$Ry$ rotations as a single native 2-qubit gate \cite{debnath2016demonstration,maslov2017basic,alexander2020qiskit}. 
	\item For Toffoli gates we use the circuit from Figure 4.8 in \cite{nielsen2010quantum}, which leads to a native 2-qubit gate count and depth of 5 and no single-qubit gates.
 	\item The comparator circuit from \cite{draper2006logarithmic}
includes $2m-2$ two-qubit gate count and $2$ two-qubit gate depth in addition to the ones
used to construct the Toffoli gates,
as well as an additional $2m+1$ single-qubit gate count and $2$ single-qubit gate depth 
aside from the ones used to construct the Toffoli gates.
\end{itemize}

\subsection{Resources for remainder loading}

The first step of the quantum (semi-)rounding methods is to load the remainder
into the amplitudes of the ancilla qubit.
The resource estimates for this step in various rounding methods are given in
Table~\ref{table:remainder_to_angle_resources} and
Table~\ref{table:remainder_to_angle_resources_2}.
We have assumed a maximum rotation bias error of $\epsilon_B = 2^{-m}$.

\begin{table}
\begin{tabular}{| l | l |  l |}
  	\hline			
  	\textbf{Resource} & Quantum Rounding 			& \\
			          & \textbf{Fault-tolerant} & \textbf{NISQ} \\
	\hline
	Total additional used qubits & $m+1$  & $m+1$ \\
	\hline
	Uncomputed ancilla qubits & $4m - \lceil\log_2{m}\rceil -2$ & $2m -
	\lceil\log_2{m}\rceil - 2$ \\
	\hline
  	T-count & $24m - 8\lceil\log_2{m}\rceil - 16$ & - \\
  	\hline
  	T-depth & $2\lceil \log_2{m}\rceil +5$ & - \\
  	\hline
  	Two-qubit gate count & $62m - 20\lceil \log_2{m}\rceil - 42$
  	                        & $32m - 10\lceil\log_2{m}\rceil - 22$ \\
  	\hline
	Two-qubit gate depth & $10\lceil \log_2{m}\rceil + 27$
	                        & $12\lceil \log_2{m}\rceil + 32$ \\
	\hline
	Single-qubit gate count & - & $32m-10\lceil \log_2{m}\rceil - 22 $ \\
  	\hline  
	Single-qubit gate depth & - & $10\lceil \log_2{m}\rceil + 27 $ \\
	\hline
\end{tabular}
\caption{The resources used to load the remainder in the ancilla amplitude
for quantum rounding given an $m$-qubit remainder register.}
\label{table:remainder_to_angle_resources}

\bigskip

\begin{tabular}{| L | L | L |}
  	\hline
  	\textbf{Resource}
			         & \text{Semi-rounding}			& \\
			         & \textbf{Fault-tolerant} & \textbf{NISQ} \\
	\hline
	\text{Total additional used qubits}
					      & $1$      & $1$ \\
	\hline
	\text{Uncomputed ancilla qubits}
						      & 2m-2 & 2m-3 \\
	\hline
  	\text{T-count}
		  & 36m + 3m\lceil1.149\log_2{\epsilon}\rceil - 12 & -\\
  	\hline
  	\text{T-depth}
		  &12m + \lceil1.149m\log_2{\epsilon}\rceil - 3 & -\\
  	\hline
  	\text{Two-qubit gate count}
		  & $20m-30$ & 12m-15 \\
  	\hline
	\text{Two-qubit gate depth}
		  & $20m-30$ & 12m-15 \\
	\hline
	\text{Single-qubit gate count}
		  & - & 2m-2 \\
  	\hline
	\text{Single-qubit gate depth}
		  & - & 2m-2 \\
	\hline
\end{tabular}
\caption{The resources used to load the remainder in the ancilla amplitude
for quantum semi-rounding given an $m$-qubit remainder register
with a maximum rotation bias error of $\epsilon$.}
\label{table:remainder_to_angle_resources_2}

\end{table}

\subsection{Resources for the controlled addition gate}
\label{subsec:ctrl_add}
The controlled addition gate in the quantum (semi-)rounding circuits
adds $\epsilon_{RD}$
to the quantum number $\ket{\bar{x}}$.
We first discuss the resources used to implement the addition of a constant
to a quantum register and
then discuss how to add the control feature.

A quantum circuit for doing an in-place addition of two $n$-bit fixed-point
numbers $a = a_{n-1}a_{n-2}...a_{0}$ and $b = b_{n-1}b_{n-2}...b_{0}$  stored
in the registers
$\ket{a}$ and $\ket{b}$ respectively can be found in Figure 5 in
\cite{draper2006logarithmic}.
However, when $a$ is a pre-determined classical constant, we can construct an
efficient circuit with the following steps.
\begin{enumerate}
	\item Construct a circuit to add 2 $n$-qubit registers $\ket{a}$ and $\ket{b}$ as described in \cite{draper2006logarithmic}.
	\item Remove all gates that are controlled on qubits $\ket{a_i}$ where $a_i=0$
	\item Remove all gates that are controlled on the now un-altered $\ket{0}$
	ancilla qubits
	\item Remove any back-to-back X, CNOT or Toffoli gates (since they are self-adjoint)
	\item Replace all the gates that are controlled on qubits $\ket{a_j}$ where $a_j=1$ with their uncontrolled versions.
	\item Remove the $\ket{a}$ quantum register.
\end{enumerate}

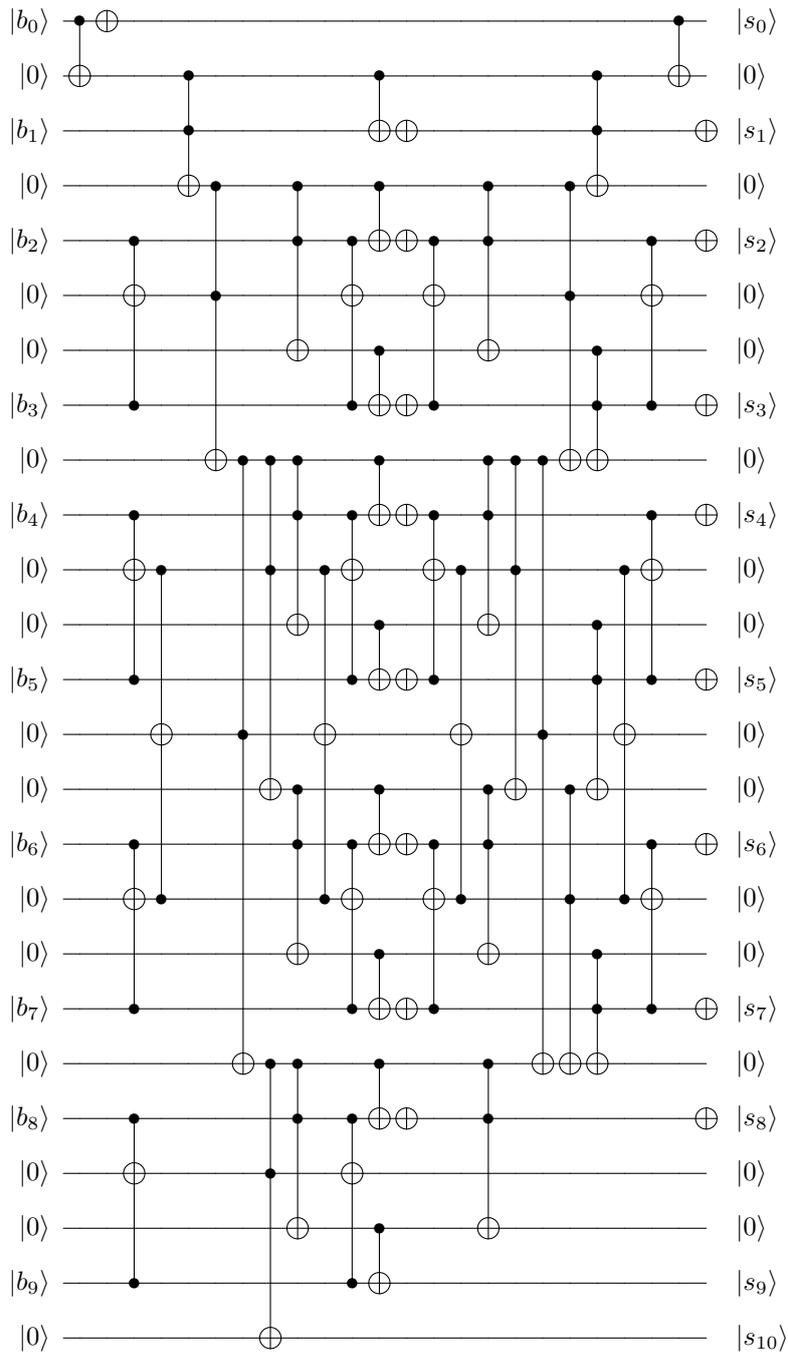
\begin{figure}
	\begin{align*}
	\Qcircuit @C=0.20em @R=1.25em {
	\lstick{\ket{b_{0}}}		 & \ctrl{1}	& \targ	& \qw      	& \qw		& \qw   	& \qw       & 
					   \qw		& \qw   	& \qw      	& \qw 		& \qw   	& \qw     & 
					   \qw		& \qw   	& \qw      	& \qw 		& \qw   	& \qw	     & 
					   \qw		& \qw   	& \qw      	& \qw 		& \ctrl{1}   	& \qw    & \rstick{\ket{s_0}} \\
	\lstick{\ket{0}}		& \targ	& \qw		& \qw      	& \qw		& \ctrl{1}   	& \qw       & 
					   \qw     	& \qw   	& \qw      	& \qw 		& \qw	 	& \ctrl{1}   &
					   \qw		& \qw   	& \qw      	& \qw 		& \qw   	& \qw	     & 
					   \qw      	& \ctrl{1}   	& \qw      	& \qw 		& \targ	& \qw       & \rstick{\ket{0}} \\
	\lstick{\ket{b_{1}}}		 & \qw		& \qw		& \qw      	& \qw		& \ctrl{1}   	& \qw       & 
					   \qw     	& \qw   	& \qw      	& \qw 		& \qw  	& \targ     &
					   \targ	& \qw   	& \qw      	& \qw 		& \qw   	& \qw       & 
					   \qw      	& \ctrl{1}  	& \qw      	& \qw 		& \qw  	& \targ      & \rstick{\ket{s_1}} \\
	\lstick{\ket{0}}		& \qw		& \qw		& \qw      	& \qw		& \targ	& \ctrl{2}  & 
					   \qw		& \qw   	& \ctrl{1}      	& \qw 		& \qw   	& \ctrl{1}    & 
					   \qw		& \qw   	& \qw      	& \ctrl{1}	& \qw   	& \qw	     & 
					   \ctrl{2}      	& \targ   	& \qw      	& \qw 		& \qw	 	& \qw        & \rstick{\ket{0}} \\
	\lstick{\ket{b_{2}}} 	& \qw		& \qw		& \ctrl{1}      	& \qw		& \qw   	& \qw       & 
					   \qw     	& \qw  	& \ctrl{2}      	&\qw		& \ctrl{1}	& \targ     &
					   \targ	& \ctrl{1}	& \qw      	& \ctrl{2}	& \qw   	& \qw       & 
					   \qw      	& \qw   	& \qw      	& \ctrl{1}	& \qw  	& \targ       & \rstick{\ket{s_2}} \\
	\lstick{\ket{0}}		& \qw		& \qw		& \targ     	& \qw		& \qw   	& \ctrl{3}   & 
					   \qw		& \qw   	& \qw      	& \qw		& \targ   	& \qw	     & 
					   \qw		& \targ   	& \qw      	& \qw 		& \qw   	& \qw	     & 
					   \ctrl{3}      	& \qw   	& \qw      	& \targ 	& \qw	 	& \qw        & \rstick{\ket{0}} \\
	\lstick{\ket{0}}		& \qw		& \qw		& \qw      	& \qw		& \qw		& \qw       & 
					   \qw		& \qw   	& \targ      	& \qw		& \qw   	& \ctrl{1} & 
					   \qw		& \qw   	& \qw      	& \targ	& \qw   	& \qw	     & 
					   \qw      	& \ctrl{1}   	& \qw      	& \qw 		& \qw	 	& \qw        & \rstick{\ket{0}} \\
	\lstick{\ket{b_{3}}}		 & \qw		& \qw		& \ctrl{-2}    	& \qw		& \qw   	& \qw       & 
					   \qw     	& \qw   	& \qw      	& \qw		& \ctrl{-2}  	& \targ     &
					   \targ	& \ctrl{-2}   	& \qw      	& \qw 		& \qw   	& \qw       & 
					   \qw      	& \ctrl{1}   	& \qw      	& \ctrl{-2}	& \qw  	& \targ       & \rstick{\ket{s_3}} \\
	\lstick{\ket{0}}		& \qw		& \qw		& \qw      	 & \qw		& \qw		& \targ     & 
					   \ctrl{5}	&\ctrl{2}	& \ctrl{1}   	& \qw      	& \qw 		& \ctrl{1}  & 
					   \qw		& \qw   	& \qw      	& \ctrl{1}	& \ctrl{2}   	& \ctrl{5}    & 
					   \targ      	& \targ   	& \qw      	& \qw 		& \qw	 	& \qw        & \rstick{\ket{0}} \\
	\lstick{\ket{b_{4}}}		& \qw		& \qw		& \ctrl{1}      	& \qw		& \qw   	& \qw       & 
					   \qw     	& \qw   	& \ctrl{2}      	& \qw		& \ctrl{1} 	& \targ     &
					   \targ	& \ctrl{1}   	& \qw      	& \ctrl{2}	& \qw   	& \qw       & 
					   \qw      	& \qw   	& \qw      	& \ctrl{1}	& \qw  	& \targ       & \rstick{\ket{s_4}} \\
	\lstick{\ket{0}}		& \qw		& \qw		& \targ      	& \ctrl{3}	& \qw   	& \qw     & 
					   \qw		& \ctrl{4}   	& \qw      	& \ctrl{3}	& \targ   	& \qw	     & 
					   \qw		& \targ   	& \ctrl{3}      	& \qw 		& \ctrl{4}   	& \qw	     & 
					   \qw      	& \qw   	& \ctrl{3}      	& \targ	& \qw	 	& \qw      & \rstick{\ket{0}} \\
	\lstick{\ket{0}}		& \qw		& \qw		& \qw      	& \qw		& \qw   	& \qw       & 
					   \qw		& \qw   	& \targ      	& \qw 		& \qw   	& \ctrl{1}   & 
					   \qw		& \qw   	& \qw      	& \targ	& \qw   	& \qw	     & 
					   \qw      	& \ctrl{1}  	& \qw      	& \qw 		& \qw	 	& \qw        & \rstick{\ket{0}} \\
	\lstick{\ket{b_{5}}} 	& \qw		& \qw		& \ctrl{-2}    	& \qw		& \qw   	& \qw       & 
					   \qw     	& \qw   	& \qw      	& \qw		& \ctrl{-2}  	& \targ     &
					   \targ	& \ctrl{-2}   	& \qw      	& \qw 		& \qw   	& \qw       & 
					   \qw      	& \ctrl{2}   	& \qw      	& \ctrl{-2}	& \qw  	& \targ       & \rstick{\ket{s_5}} \\
	\lstick{\ket{0}}		& \qw		& \qw		& \qw      	& \targ	& \qw   	& \qw    & 
					   \ctrl{6}	& \qw   	& \qw      	& \targ	& \qw   	& \qw	     & 
					   \qw		& \qw   	& \targ      	& \qw 		& \qw   	& \ctrl{6}    & 
					   \qw      	& \qw   	& \targ      	& \qw 		& \qw	 	& \qw        & \rstick{\ket{0}} \\
	\lstick{\ket{0}}		& \qw		& \qw		& \qw      	& \qw 		& \qw		& \qw	&
					   \qw		& \targ    	& \ctrl{1}      	& \qw 		& \qw  	& \ctrl{1}  & 
					   \qw		& \qw   	& \qw      	& \ctrl{1}	& \targ   	& \qw	     & 
					   \ctrl{2}     	& \targ   	& \qw      	& \qw 		& \qw	 	& \qw        & \rstick{\ket{0}} \\
	\lstick{\ket{b_{6}}}		 & \qw		& \qw		& \ctrl{1}      	& \qw		& \qw   	& \qw       & 
					   \qw     	& \qw   	& \ctrl{2}     	& \qw		& \ctrl{1}	& \targ    &
					   \targ	& \ctrl{1}  	& \qw      	& \ctrl{2}	& \qw   	& \qw       & 
					   \qw      	& \qw   	& \qw      	& \ctrl{1}	& \qw  	& \targ       & \rstick{\ket{s_6}} \\
	\lstick{\ket{0}}		& \qw		& \qw		& \targ      	& \ctrl{-3}	& \qw  	& \qw       & 
					   \qw		& \qw   	& \qw	    	& \ctrl{-3}    & \targ   	& \qw	     & 
					   \qw		& \targ   	& \ctrl{-3}    	& \qw 		& \qw   	& \qw	     & 
					   \ctrl{3}      	& \qw   	& \ctrl{-3}   	& \targ	& \qw	 	& \qw        & \rstick{\ket{0}} \\
	\lstick{\ket{0}}		& \qw		& \qw		& \qw      	& \qw		& \qw	   	& \qw       & 
					   \qw		& \qw   	& \targ      	& \qw 		& \qw   	& \ctrl{1}    & 
					   \qw		& \qw   	& \qw      	& \targ	& \qw   	& \qw	     & 
					   \qw      	& \ctrl{1}   	& \qw      	& \qw 		& \qw	 	& \qw        & \rstick{\ket{0}} \\
	\lstick{\ket{b_{7}}} 	& \qw		& \qw		& \ctrl{-2}    	& \qw		& \qw   	& \qw       & 
					   \qw     	& \qw   	& \qw      	& \qw		&  \ctrl{-2}	& \targ     &
					   \targ	& \ctrl{-2}   	& \qw      	& \qw 		& \qw   	& \qw       & 
					   \qw      	& \ctrl{1}   	& \qw      	& \ctrl{-2}	& \qw  	& \targ       & \rstick{\ket{s_7}} \\
	\lstick{\ket{0}}		& \qw		& \qw		& \qw      	& \qw		& \qw	   	& \qw      & 
					   \targ	& \ctrl{2}	& \ctrl{1}   	& \qw      	& \qw 		& \ctrl{1}     & 
					   \qw		& \qw   	& \qw      	& \ctrl{1}	& \qw   	& \targ     & 
					   \targ      	& \targ   	& \qw      	& \qw 		& \qw	 	& \qw        & \rstick{\ket{0}} \\
	\lstick{\ket{b_{8}}} 	& \qw		& \qw		& \ctrl{1}      	& \qw		& \qw   	& \qw       & 
					   \qw     	&  \qw  	& \ctrl{2}     	& \qw		& \ctrl{1}  	& \targ     &
					   \targ	& \qw   	& \qw      	& \ctrl{2}	& \qw   	& \qw       & 
					   \qw      	& \qw   	& \qw      	& \qw 		& \qw  	& \targ       & \rstick{\ket{s_8}} \\
	\lstick{\ket{0}}		& \qw		& \qw		& \targ      	& \qw		& \qw   	& \qw       & 
					   \qw		& \ctrl{3}   	& \qw      	& \qw		& \targ   	& \qw	     & 
					   \qw		& \qw   	& \qw      	& \qw 		& \qw   	& \qw	     & 
					   \qw      	& \qw   	& \qw      	& \qw 		& \qw	 	& \qw        & \rstick{\ket{0}} \\
	\lstick{\ket{0}}		& \qw		& \qw		& \qw      	& \qw		& \qw   	& \qw       & 
					   \qw		& \qw   	& \targ      	& \qw 		& \qw   	& \ctrl{1}    & 
					   \qw		& \qw   	& \qw    	& \targ	& \qw   	& \qw	     & 
					   \qw      	& \qw   	& \qw      	& \qw 		& \qw	 	& \qw        & \rstick{\ket{0}} \\
	\lstick{\ket{b_{9}}}		 & \qw		& \qw		& \ctrl{-2}    	& \qw		& \qw   	& \qw       & 
					   \qw     	& \qw   	& \qw      	& \qw		& \ctrl{-2}  	& \targ       &
					   \qw		& \qw   	& \qw      	& \qw 		& \qw   	& \qw       & 
					   \qw      	& \qw   	& \qw      	& \qw 		& \qw  	& \qw       & \rstick{\ket{s_9}} \\
	\lstick{\ket{0}}		& \qw		& \qw		& \qw      	& \qw		& \qw   	& \qw       & 
					   \qw		& \targ   	& \qw      	& \qw 		& \qw   	& \qw	     & 
					   \qw		& \qw   	& \qw      	& \qw 		& \qw   	& \qw	     & 
					   \qw      	& \qw   	& \qw      	& \qw 		& \qw	 	& \qw        & \rstick{\ket{s_{10}}}
	}
	\end{align*}
	\caption{A circuit to add the constant $\epsilon_{RD}$ to a 10-qubit
	register. The circuit was constructed by taking the the addition circuit
	from Figure 5 in \cite{draper2006logarithmic} and performing the relevant
	modifications as described in Section~\ref{subsec:ctrl_add}.}
	\label{fig:add_1}
\end{figure}

For our particular circuits we want to implement the $ADD_{\epsilon_{RD}}$ gate where we replace
 $a_0=1$ and $a_i=0$ for $i \neq 0$.
This leads to a further reduction in resources used to construct
this gate compared to
the resources used to add 2 arbitrary $n$ qubit registers as described in
Section 4.2 in \cite{draper2006logarithmic}. The T-count is reduced by 35\%
while the number of ancilla qubits remains the same. An example of a circuit to
add 1 to a $10$-qubit register can be seen in \autoref{fig:add_1}.

\begin{figure}
	\begin{align*}
	\Qcircuit @C=0.20em @R=1.25em {
	\lstick{\ket{c}} & \ctrl{1}  	& &     & & & \qw & \ctrl{1}   & \qw\\
	\lstick{\ket{a_0}} & \qswap 		& &
	\push{\rule{.3em}{0em}=\rule{.3em}{0em}}
	 & & &
	\targ & \ctrl{1}       & \targ\\
	\lstick{\ket{a_1}} & \qswap \qwx	& &	& & & \ctrl{-1}    &  \targ & \ctrl{-1}
	}
	\end{align*}
	\caption{The decomposition of a controlled-swap gate into Toffoli and CNOT gates.}
	\label{fig:ctrl_swap}
\end{figure}
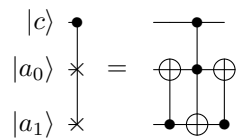

Once we have constructed the $ADD_{\epsilon_{RD}}$ gate, we can turn it into a
controlled-$ADD_{\epsilon_{RD}}$ gate by using the method shown in Figure 3 in
\cite{maslov2011reversible}. This uses an additional $n$-qubit ancilla
register, along with two sets of controlled swap gates. Each individual
controlled swap gate comprises 3 Toffoli gates as seen in
\autoref{fig:ctrl_swap}. The resources used for a controlled
$ADD_{\epsilon_{RD}}$ gate using this procedure are given in
Table~\ref{table:ctrl_add_1} and \autoref{table:ctrl_add_1_2}.

\begin{table}
\begin{tabular}{| l | l |}
  	\hline			
  	\textbf{Resource} & \textbf{Fault-tolerant implementation} \\
	\hline
	Total additional used qubits & $1$ \\
	\hline
	Uncomputed ancilla qubits & $3n - \lfloor\log_2{n}\rfloor+
	        \sum_{i=1}^{\log_2{n}}\lfloor\frac{n}{2^i}\rfloor - 5$ \\
	\hline
  	T-count & $34n - 12\lfloor\log_2{n}\rfloor -
        12\lfloor\log_2{n-1}\rfloor$ \\
        & \qquad $+ 12\sum_{i=1}^{\log_2{n}}\lfloor\frac{n}{2^i}\rfloor$ \\
        & \qquad $+ 12\sum_{i=1}^{\log_2{n-1}}\lfloor\frac{n-}{2^i}\rfloor - 12
         $ \\
  	\hline
	T-depth & $\lfloor\log_2{n}\rfloor + \lfloor\log_2{n-1}\rfloor $ \\
	& \qquad $ + \lfloor\log_2{\frac{n}{3}}\rfloor + \lfloor\log_2{\frac{n-1}{3}}\rfloor + 11$ \\
  	\hline
  	Two-qubit gate count & $77n - 30\lfloor\log_2{n}\rfloor -
  	30\lfloor\log_2{n-1}\rfloor $ \\
  	& \qquad $ + 30\sum_{i=1}^{\log_2{n}}\lfloor\frac{n}{2^i}\rfloor $ \\
  	& \qquad $ + 30\sum_{i=1}^{\log_2{n-1}}\lfloor\frac{n-1}{2^i}\rfloor - 29 $
  	 \\
  	\hline
	Two-qubit gate depth & $10\lfloor\log_2{n}\rfloor + 10\lfloor\log_2{n-1}\rfloor $ \\
	& \qquad $+ 10\lfloor\log_2{\frac{n}{3}}\rfloor + 10\lfloor\log_2{\frac{n-1}{3}}\rfloor + 111$ \\
	\hline
\end{tabular}
\caption{The resources used to perform a ctrl-add 1 to an $n$-qubit
register using the fault-tolerant implementation.}
\label{table:ctrl_add_1}

\bigskip

\begin{tabular}{| l | l |}
  	\hline
  	\textbf{Resource} & \textbf{NISQ implementation} \\
	\hline
	Total additional used qubits
		     			  & $1$ \\
	\hline
	Number of uncomputed ancilla qubits & $n - \lfloor\log_2{n}\rfloor+
        \sum_{i=1}^{\log_2{n}}\lfloor\frac{n}{2^i}\rfloor - 1$ \\
	\hline
  	Two-qubit gate count & $\lceil44.2n\rceil - 15\lfloor\log_2{n}\rfloor - 15\lfloor\log_2{n-1}\rfloor$
  	 \\
               & \qquad $+ 15\sum_{i=1}^{\log_2{n}}\lfloor\frac{n}{2^i}\rfloor
               $ \\
               & \qquad $+ 15\sum_{i=1}^{\log_2{n-1}}\lfloor\frac{n-1}{2^i}\rfloor -
               14$
         \\
  	\hline
	Two-qubit gate depth & $5\lfloor\log_2{n}\rfloor + 5\lfloor\log_2{n-1}\rfloor $ \\
	           & \qquad $+ 5\lfloor\log_2{\frac{n}{3}}\rfloor $ \\
	           & \qquad $+ 5\lfloor\log_2{\frac{n-1}{3}}\rfloor + 56$ \\
	\hline
	Single-qubit gate count
		   		     &  $ 2n+1 $ \\
  	\hline
	Single-qubit gate depth
		   		     & $2$ \\
	\hline
\end{tabular}
\caption{The resources used to perform a ctrl-add 1 to an $n$-qubit
register using the NISQ implementation.}
\label{table:ctrl_add_1_2}

\end{table}

\section{Algorithm and resources for generalized fixed-point multiplication}
\label{sec:fpm}
In this section we consider the resources used to perform a fixed-point
multiplication of two numbers $a$ and $b$. We assume that they
can be represented by the binary numbers
\begin{align*}
	a &= a_{p_a}a_{p_a-1}...a_0.a_{-1}...a_{p_a-n_a} \\
	b &= b_{p_b}b_{p_b-1}...b_0.b_{-1}...b_{p_b-n_b}
\end{align*}
and we will store them in two seperate registers $\ket{a}$ and $\ket{b}$.
In Appendix A in \cite{haner2018optimizing}, the authors state that fixed point
multiplication can be computed using a series of controlled
additions e.g. initialize an output register $\ket{00...0}$ and perform $n_a$
ctrl-additions of $2^{j} \times b$ for $j$ ranging from $p-n$ to $p$
where each $j$th addition is controlled on $a_j$. In Appendix B of
\cite{haner2018optimizing} the authors describe the T-count for the case where the
two input registers and the output register have the same values for $n$ and $p$, leading to the final result having a maximum error of $\frac{n}{2^{n-p}}$.
We extend this original algorithm to be for any 
two distinct input register sizes and as a function of a desired error
$\epsilon$ leading to an output register with size parameters $n_{out}$ and $p_{out}$.
We do this by first recognizing that the multiplication includes $n_a$
separate control-additions to the output register e.g. one control-addition for
each $a_j$ term. The resources used to perform a control-addition are
discussed in Section~\ref{subsec:ctrl_add}. Naively we would use all the
$n_b$ bits in $b$ for each addition. However, for a given error, we can use
less than $n_b$ bits in some of these  additions. To compute this number, we
start by noticing
that given a total error of $\epsilon$, this means that we can tolerate
an error of up to $\frac{\epsilon}{n_a}$ for each controlled-addition. 
Each controlled-addition term will have the value $a_j\times 2^{j}$ times $b$.
Therefore we may only use the $f_j$ most significant ones
for every $a_j$ term, where $f_j$ is computed by solving for $2^{p_b-f_j} \leq \frac{\epsilon}{n_a}$. For values of $j$ for which $f_j > n_b$, we use all $n_b$ bits.
We note that the final output register size parameters will be
\begin{align*}
	p_{out} &= p_a + p_b \\
	n_{out} &= p_{out} - \log_2{\epsilon}
\end{align*}

We finally note that the case of the Haner et al. \cite{haner2018optimizing} method is the special case where $\epsilon = \frac{n}{2^{n-p}}$, 
and the Exact method mentioned in the text (i.e. $\epsilon=0$) is also a special case 
in which we end up performing $n_a$ control-additions all with $n_b$ bits registers.

\end{document}